\documentclass{mn2e}

\usepackage{times}
\usepackage{epsfig} 
\usepackage{amssymb}
\usepackage{graphics}
\usepackage{natbib}

\newcommand{\ergs}{erg~s$^{-1}$} 
\newcommand{\nh}{$N_{\rm H}$} 
\newcommand{\ch}{{\it Chandra}}
\newcommand{\xmmn}{{\it XMM-Newton}}

\newcommand{\spz}{{\it Spitzer}}
\newcommand{\her}{{\it Herschel}} 
 
\begin{document}
\title[Infrared selected heavily-obscured quasars at $z\approx2$]{Mid-infrared luminous quasars in the GOODS-\her\ fields: a large population of heavily-obscured, Compton-thick quasars at $z\approx2$}

\author[A.~Del Moro et al.]{A.~Del Moro,$^{1,2}$
D.~M.~Alexander,$^1$ F.~E.~Bauer,$^{3,4,5,6}$ E. Daddi,$^7$ D.~D.~Kocevski,$^8$ \and
D.~H.~McIntosh,$^9$ F. Stanley,$^1$ W.~N.~Brandt,$^{10,11,12}$ D.~Elbaz,$^7$ C.~M.~Harrison,$^1$ \and
B.~Luo,$^{13}$ J.~R.~Mullaney,$^{14}$ and Y.~Q.~Xue$^{15}$ \\ 
$^1$ Centre for Extragalactic Astronomy, Department of Physics, Durham University, South Road, Durham, DH1 3LE, UK\\
$^2$ Max-Planck-Institut f{\"u}r Extraterrestrische Physik (MPE), Postfach 1312, D85741, Garching, Germany \\
$^3$ Instituto de Astrof\'{\i}sica, Facultad de F\'{i}sica, Pontificia Universidad Cat\'{o}lica de Chile, 306, Santiago 22, Chile\\
$^4$ Millenium Institute of Astrophysics, Santiago, Chile\\
$^5$ EMBIGGEN Anillo, Concepci\'{o}n, Chile\\
$^6$ Space Science Institute, 4750 Walnut Street, Suite 205, Boulder, Colorado 80301\\
$^7$ CEA-Saclay, 91191 Gif-sur-Yvette Cedex, France\\
$^8$ Department of Physics and Astronomy, University of Kentucky, Lexington, KY 40506-0055, USA \\
$^9$ Department of Physics \& Astronomy, University of Missouri-Kansas City, 5110 Rockhill Rd., Kansas City, MO 64110, USA\\
$^{10}$ Department of Astronomy and Astrophysics, 525 Davey Lab, Pennsylvania State University, University Park, PA 16802, USA\\
$^{11}$ Institute for Gravitation and the Cosmos, Pennsylvania State University, University Park, PA 16802, USA \\
$^{12}$ Department of Physics, Pennsylvania State University, University Park, PA 16802, USA \\
$^{13}$ School of Astronomy and Space Science, Nanjing University, Nanjing 210093, China \\
$^{14}$ Department of Physics and Astronomy, University of Sheffield, Hounsfield Road, Sheffield S3 7RH, UK\\
$^{15}$ Key Laboratory for Research in Galaxies and Cosmology, Department of Astronomy, University of Science and Technology of China, \\
Chinese Academy of Sciences, Hefei, Anhui 230026, China}
\date{}
\maketitle

\begin{abstract}
We present the infrared (IR) and X-ray properties of a sample of 33 mid-IR luminous quasars ($\nu L_{6 \mu m}\ge6\times10^{44}$ erg~s$^{-1}$) at redshift $z\approx$~1--3, identified through detailed spectral energy distribution analyses of distant star-forming galaxies, using the deepest IR data from \spz\ and \her\ in the GOODS-\her\ fields. The aim is to constrain the fraction of obscured, and Compton-thick (CT, \nh$>1.5\times10^{24}$~cm$^{-2}$) quasars at the peak era of nuclear and star-formation activities.
Despite being very bright in the mid-IR band, $\approx$30\% of these quasars are not detected in the extremely deep 2 Ms and 4 Ms \ch\ X-ray data available in these fields. X-ray spectral analysis of the detected sources reveals that the majority ($\approx$67\%) are obscured by column densities \nh$>10^{22}$~cm$^{-2}$; this fraction reaches $\approx$80\% when including the X-ray undetected sources (9 out of 33), which are likely to be the most heavily-obscured, CT quasars. We constrain the fraction of CT quasars in our sample to be $\approx$24-48\%, and their space density to be $\Phi=({6.7}\pm2.2)\times10^{-6}$ Mpc$^{-3}$. 
From the investigation of the quasar host galaxies in terms of star-formation rates (SFRs) and morphological distortions, as a sign of galaxy mergers/interactions, we do not find any direct relation between SFRs and quasar luminosity or X-ray obscuration. On the other hand, there is tentative evidence that the most heavily-obscured quasars have, on average, more disturbed morphologies than the unobscured/moderately-obscured quasar hosts, which preferentially live in undisturbed systems. However, the fraction of quasars with disturbed morphology amongst the whole sample is $\approx$40\%, suggesting that galaxy mergers are not the main fuelling mechanism of quasars at $z\approx2$. 
\end{abstract}



\section{Introduction}\label{intro}
Since the discovery that most, if not all, galaxy spheroids in the local Universe host a supermassive black hole (SMBH) whose mass scales with that of the host-galaxy bulge (e.g., \citealt{magorrian1998,ferrarese2000,marconi2003}) it has been clear that there is likely to be a tight connection between the evolution of SMBHs and galaxies. Studies of the galaxy population across cosmic time have shown that galaxies were more actively growing in the past, as the star-formation rate (SFR) density peaks at redshift $z\approx2$ and rapidly declines to the present day (e.g., \citealt{madau1996,hopkins2006a}). A similar evolution is seen in the moderate-to-high luminosity active galactic nucleus (AGN) population, where most of the SMBHs accrete more rapidly at high redshift ($z\gtrsim$~1-2), becoming more quiescent at $z=0$ (e.g., \citealt{brandt2015}, and references therein). Whether this parallel evolution between galaxies and SMBHs is simply due to a larger gas supply to feed both black hole accretion and star formation at high redshift, or whether there are other processes that self-regulate the SMBH and galaxy growth is still uncertain (e.g., \citealt{alexander2012,kormendy2013}). A complete census of AGN is crucial to fully understand the accretion history of the SMBHs and their role in galaxy evolution. However, as the Universe at high redshift is more gas and dust rich, most of the accretion onto SMBHs is expected to be heavily obscured, which makes the identification and study of a large fraction of the AGN population very challenging.

X-ray observations provide arguably the most efficient way to trace the growth of SMBHs, as the high energy radiation can penetrate large amounts of gas and dust without significant absorption. Moreover, the contamination from galaxy emission at X-ray energies is small as galaxies are typically much fainter than  AGN in this band. However, when the column density of the obscuring material becomes Compton thick (CT; $N_{\rm H}>1.5\times10^{24}$~cm$^{-2}$) even the deepest X-ray surveys often fail to detect the population of the most obscured, CT AGN (e.g., \citealt{xue2012}; see \citealt{brandt2015}, for a review). A significant fraction of such AGN is predicted by all models of the Cosmic X-ray background (CXB) to reproduce the observed background spectrum at high energies ($E\approx30$~keV; e.g., \citealt{comastri1995,gilli2007,treister2009}), however, the actual contribution from CT AGN is not well constrained and varies between the models, due to different assumptions, and the space density of CT AGN is still uncertain. 

Since most of the absorbed AGN emission is re-emitted in the mid-infrared band (MIR; $\lambda\approx$~3-40~$\mu$m) by the circumnuclear dust heated by the central SMBH, observations in this band can potentially identify the elusive CT AGN missed by X-ray surveys (e.g., \citealt{lacy2004, stern2005, fiore2008, alexander2008b,georgantopoulos2011b, donley2008,donley2012,mateos2013}). Moreover, since the infrared (IR) band is much less affected by extinction than the UV, optical and soft X-ray bands, the selection of AGN in the MIR band is almost unbiased against obscured AGN. The downside of using the IR band to study AGN is the strong contamination from star formation (SF) in the host galaxies, whose cold dust emission tends to dominate the MIR to far-IR (FIR, $\lambda\approx$~40-250~$\mu$m) bands, peaking at $\lambda\approx$~100-160~$\mu$m. This is especially the case for low-to-moderate luminosity AGN, whose IR emission is often swamped by SF. These sources are easily missed by typical MIR colour-colour sample selection, which are generally used to identify AGN within the star-forming galaxy (SFG) population (e.g., \citealt{lacy2004, stern2005,donley2012,mateos2013}).

In this paper we investigate a sample of MIR luminous quasars (hereafter, ``IR quasars'') at redshift $z\approx$~1-3, i.e., at the peak era of AGN and SF activity. We used the deepest available \spz\ and \her\ data in the GOODS-\her\ North and South fields (hereafter, GH-N and GH-S, respectively; \citealt{elbaz2011}) to reliably identify AGN in the MIR band through detailed spectral energy distribution (SED) analyses, which allow us to disentangle the AGN and SF emissions at MIR and FIR wavelengths. The SED analysis is a much more powerful tool than the simple IR colour-colour techniques for identifying AGN, as it can recover the hot dust emission from the AGN even when it does not strongly dominate the SED (e.g., \citealt{delmoro2013,rovilos2014}); however even the SED analyses have limitations and become incomplete when the AGN component is faint compared to the SF component. Using the deepest \ch\ X-ray data available across the sky (2 Ms in CDF-N; \citealt{alexander2003}; and 4 Ms in CDF-S; \citealt{xue2011}), we investigate the X-ray spectra of the IR quasars to constrain the fraction of obscured and CT sources. We also investigate the properties of their host galaxies to place them in the context of potential SMBH-galaxy evolution scenarios.
The paper is organised as follows: in Section 2 the data used in our analyses are described, while Section 3 details the SED analysis method and the selection of our IR quasar sample. Section 4 presents the X-ray spectral analysis and results for our quasars, as well as the properties of their host galaxies assessed by the SFRs derived from our SED analyses, and their morphologies, obtained from visual classification of $HST$-WFC3 $K$-band images. In Section 5 the results obtained from our analyses are discussed in the wider context of the CXB models and the SMBH-galaxy evolutionary scenario. A summary and final remarks are given in Section 6.
Throughout the paper we assume a cosmological model with $H_0=$~70~$\rm km~s^{-1}~Mpc^{-1}$, $\Omega_{\rm M}=0.27$ and $\Omega_{\Lambda}=0.73$ \citep{spergel2003}. All the errors are quoted at a 90\% confidence level, unless otherwise stated.

\section{Data}\label{data}

\subsection{Infrared Observations}\label{ir}
The GOODS fields have been observed in the infrared band by the \spz\ and \her\ space observatories, and include the deepest IR data currently available. 
The \spz\ observations \citep{giavalisco2004} cover the MIR band at 3.6, 4.5, 5.8 and 8.0 $\mu$m with IRAC (\citealt{fazio2004}) and at 24~$\mu$m with MIPS (\citealt{rieke2004}) as part of the GOODS Legacy program (PI: M. Dickinson). The 24~$\mu$m source catalogue was created using a PSF fitting technique at the 3.6~$\mu$m source positions as priors ($>5\sigma$ detections). For details of the observations and catalogue we refer to \citet{magnelli2009} and \citet{magnelli2011}. In both fields the 24~$\mu$m observations reach a point source sensitivity limit of $\sim30\ \mu$Jy (5$\sigma$).

FIR observations of the GOODS fields were undertaken by \her\ as part of the GOODS-\her\ survey project (GH; \citealt{elbaz2011}). The GH survey covers the full $\approx$10$'\times$16$'$ GOODS-N field (hereafter, GH-N) with PACS (\citealt{poglitsh2008}) at 100 and 160 $\mu$m and with SPIRE (\citealt{griffin2010}) at 250, 350, 500 $\mu$m, for a total observing time of 124 h. In GOODS-S a smaller region of $\approx$10$'\times$10$'$ at the centre of the field was observed by PACS at 100 and 160 $\mu$m and the total exposure was 206.3 h (GH-S; \citealt{elbaz2011}). We also used the SPIRE 250 $\mu$m data observed as part of the \her\ Multi-tiered Extragalactic Survey (HerMES; \citealt{oliver2012}) across the GH-S field, to better match the data available in GH-N. We note that since the SPIRE data suffer from significant flux blending issues, we limit our analyses to the 250~$\mu$m band in both the GH-N and GH-S fields, where these issues can still be corrected fairly well.  

The \spz-MIPS 24~$\mu$m source positions were used as priors to estimate the \her\ 100 $\mu$m, 160 $\mu$m and 250 $\mu$m fluxes through PSF fitting (courtesy of E. Daddi). In the GH-N field 819 sources ($\sim$42\% of the 24~$\mu$m priors) are detected in at least one of the \her\ bands down to 5$\sigma$ (3$\sigma$) sensitivity limits of $\sim$1.7 ($\sim$1.2) mJy, $\sim$4.5 ($\sim$2.3) mJy and $\sim$6.5 ($\sim$4.0) mJy, respectively. In the GH-S field there are 591 sources ($\sim$34\% of the 24~$\mu$m priors) detected by \her\ (at 100 $\mu$m, 160 $\mu$m or 250 $\mu$m); the faintest sources are detected down to 5$\sigma$ (3$\sigma$) flux densities of 1.3 (0.7) mJy, 2.7 (1.7) mJy, 7.7 (5.0) mJy in the three bands, respectively; however, for general sources the flux densities and errors are higher, depending on the local source density and the de-blending uncertainties. In case of non-detection in any of the \her\ bands, appropriate flux density upper limits were calculated for each prior position. 

In our analyses we initially considered all the 24~$\mu$m-detected sources within the GH fields, namely 1943 sources in GH-N and 1747 sources in GH-S (for a total of 3690 sources), with or without a \her\ detection. For the sample in GH-N $\approx$98\% of the sources have a redshift identification, of which $\approx$63\% are spectroscopic redshifts, while the remaining are photometric (see \citealt{delmoro2013}, hereafter DM13, for a detailed description of the redshift compilation). In the GH-S field, a redshift identification is available for $\approx$86\% of the sources, of which $\approx$95\% are spectroscopic redshifts compiled from multiple catalogues available in the literature (\citealt{lefevre2004, mignoli2005, vanzella2008, popesso2009, balestra2010, silverman2010, xia2011}) and 5\% are photometric redshifts.

\subsection{X-ray Observations}\label{xdata}
The X-ray counterparts for the 24~$\mu$m detected sources in GH-N and GH-S were found by cross-matching the 24~$\mu$m positions with the X-ray source catalogues by \citet{alexander2003} in the \ch\ Deep Field North (CDF-N), constructed from the 2 Ms \ch\ observations covering an area of $\sim$448 arcmin$^2$ of the sky, and by \citet{xue2011} in the \ch\ Deep Field South (CDF-S), from the 4 Ms \ch\ data over a region of 464.5 arcmin$^2$.
The CDF-N catalogue contains 503 X-ray detected sources, down to flux limits of $f_{\rm 0.5-2\ keV}\approx2.5\times10^{-17}\ \rm erg~cm^{-2}~s^{-1}$  and $f_{\rm 2-8\ keV}\approx1.4\times10^{-16}\ \rm erg~cm^{-2}~s^{-1}$. The CDF-S X-ray source catalogue contains 740 sources down to flux limits of $f_{\rm 0.5-2\ keV}\approx9.1\times10^{-18}\ \rm erg~cm^{-2}~s^{-1}$ and $f_{\rm 2-8\ keV}\approx5.5\times10^{-17}\ \rm erg~cm^{-2}~s^{-1}$. For details on the data reduction and the construction of the catalogues we refer to \citet{alexander2003}, \citet{luo2008} and \citet{xue2011}. 

For the X-ray detected sources in our sample (see Sect. \ref{irquasar}), X-ray spectra have been extracted and analysed. The data were processed using the \ch\ Interactive Analysis of Observations\footnote{http://cxc.cfa.harvard.edu/ciao/index.html} (CIAO; version 4.3 and CALDB 4.4.1.; \citealt{fruscione2006}) tools and the {\em ACIS Extract} (AE) software package\footnote{The {\em ACIS Extract} software package and Users Guide are available at http://www.astro.psu.edu/xray/acis/acis\_analysis.html.} \citep{broos2010,broos2012}. 
The \ch\ spectra were produced using the AE software (version 2011-03-16), extracting source and background spectra in each individual observation, as well as the relative response matrices and ancillary files, and combining them using the FTOOLS \texttt{addrmf} and \texttt{addarf}. 

\subsection{Optical/Near-Infrared Observations}\label{odata}

The GOODS-N and GOODS-S fields have been observed by the {\it Hubble} Space telescope (HST) ACS camera as part of the GOODS project \citep{giavalisco2004}, and WFC3 camera as part of the Cosmic Near-Infrared Deep Extragalactic Legacy Survey (CANDELS), a large program consisting of broad-band photometric UV to the near-IR imaging of five separate deep extragalactic fields (see, \citealt{grogin2011} and \citealt{cocomero2011}, for details on the observations and data reduction). The CANDELS HST observations in the GOODS-N field cover an area of $\approx10'\times16'$ (CANDELS-Wide) with a deeper smaller region in the centre (CANDELS-DEEP; $\approx6.8'\times10'$). The imaging of the GOODS-S field covers a similar region as in GOODS-N, but also includes a small ultra-deep area in the centre of the field ({\it Hubble} Ultra-Deep field; $2'\times2.3'$). In this paper we mainly use data in the F125W and F160W bands, which approximately correspond to near-IR $J$ and $H$ bands, but also consider data from the optical F606W and F850LP bands (corresponding to the $V$ and $z$ bands, respectively), to corroborate our visual classification of the galaxy morphology (Section \ref{morph}).

\section{Spectral Energy Distribution and sample selection}
\subsection{Infrared SED decomposition}\label{sed}

To constrain the IR emission from the AGN and the host galaxies, our SED fitting approach uses an empirical AGN template and five different star-forming galaxy templates from \citet{mullaney2011}, which have been extended down to 3 $\mu$m and to the radio band (DM13). The AGN template at $\lambda\lesssim40\ \mu$m is represented by a broken power law, where the indices have been fixed at the average values of $\Gamma_1=1.7$ and $\Gamma_2=0.7$ and the power-law break at $\lambda=19\ \mu$m.\footnote{We also performed the SED fitting using $\Gamma_1$ as a free parameter for the sources with enough data points in the MIR band to constrain the power-law slope. We note that this is not the case for all the sources, as at $z\gtrsim1.7$ the 8~$\mu$m data point falls out of the wavelength range covered by our SED templates. Moreover, in some cases the 16~$\mu$m data are upper limits and therefore cannot provide strong constraints to the MIR power-law slope. However, where the fits with $\Gamma_1$ as free parameter provided a better representation of the data compared to a fixed $\Gamma_1=1.7$, we adopted these fits as the best-fitting solutions.} An extinction law (\citealt{draine2003}) is also applied to the AGN template, with the extinction parameter $A_{\rm V}$ free to vary in the range of $A_{\rm V}=$~0-30~mag (\citealt{mullaney2011}; DM13). The \spz\ 8, 16 and 24~$\mu$m and the \her\ 100, 160 and 250~$\mu$m photometric points are used to constrain the source SEDs (see Figures \ref{fig.sed} and \ref{fig.sed1} in Appendix); the \spz\ IRAC flux densities at 3.6, 4.5, and 5.8~$\mu$m are not included in the SED analysis because these data are likely to be dominated by starlight emission, which is not taken into account in our templates. 

The SEDs for all of the 24~$\mu$m-detected sources with redshift measurements in the GH-N and GH-S fields have been initially fitted with the SFG templates only, and subsequently with the AGN $+$ SFG templates obtaining a best-fitting solution for each of the SFG templates, using a $\chi^2$ fitting technique (see DM13, for details). We adopted the Bayesian information criterion (BIC; \citealt{schwarz1978}) to assess the improvement of the fits due to the addition of the AGN component to each of the SFG templates, and therefore to identify the best-fitting model for each source. Namely, the AGN $+$ SFG model is assumed as the best-fitting model if for each source: 
\begin{itemize}
\item[(1)] $\Delta \rm BIC_i=BIC_{SFG,i}-BIC_{AGN+SFG,i}>2$ (e.g., \citealt{kass1995}), where $i=$~1-5 refers to each of the SFG templates, in at least 4 out of the 5 pairs of best-fitting solutions; \\
\item[(2)] $\rm min(\chi^2_{AGN+SFG,j})<min(\chi^2_{SFG,i})$, with $j\le\ i$, i.e., the minimum $\chi^2$ of the AGN $+$ SFG solutions that satisfy condition (1) should be less than the minimum $\chi^2$ obtained by fitting any of the SFG templates alone.  
\end{itemize}
If any of these criteria is not met we conservatively adopt the simple SFG model as the best-fitting model. 
These criteria are adopted to assess the reliability of the AGN component independently from the individual SFG templates. In fact, due to the sparse data points available for our SED analysis, in some cases, individual fitting solutions from different templates yield similarly good fit to the data, preventing us from identifying a unique best-fitting solution. 
We refer to Section 3.1 and Appendix A of DM13, for a detailed description of the SED fitting approach and tests on its reliability.\footnote{We note that here we have slightly modified the criteria to evaluate the reliability of the AGN component compared to those presented in DM13, by adopting the BIC instead of an $f$-test. The BIC is a more appropriate statistical criterion to compare the results from different models than the $f$-test when the models are non-nested.}  
 
\begin{figure}
\centerline{
\includegraphics[scale=0.58]{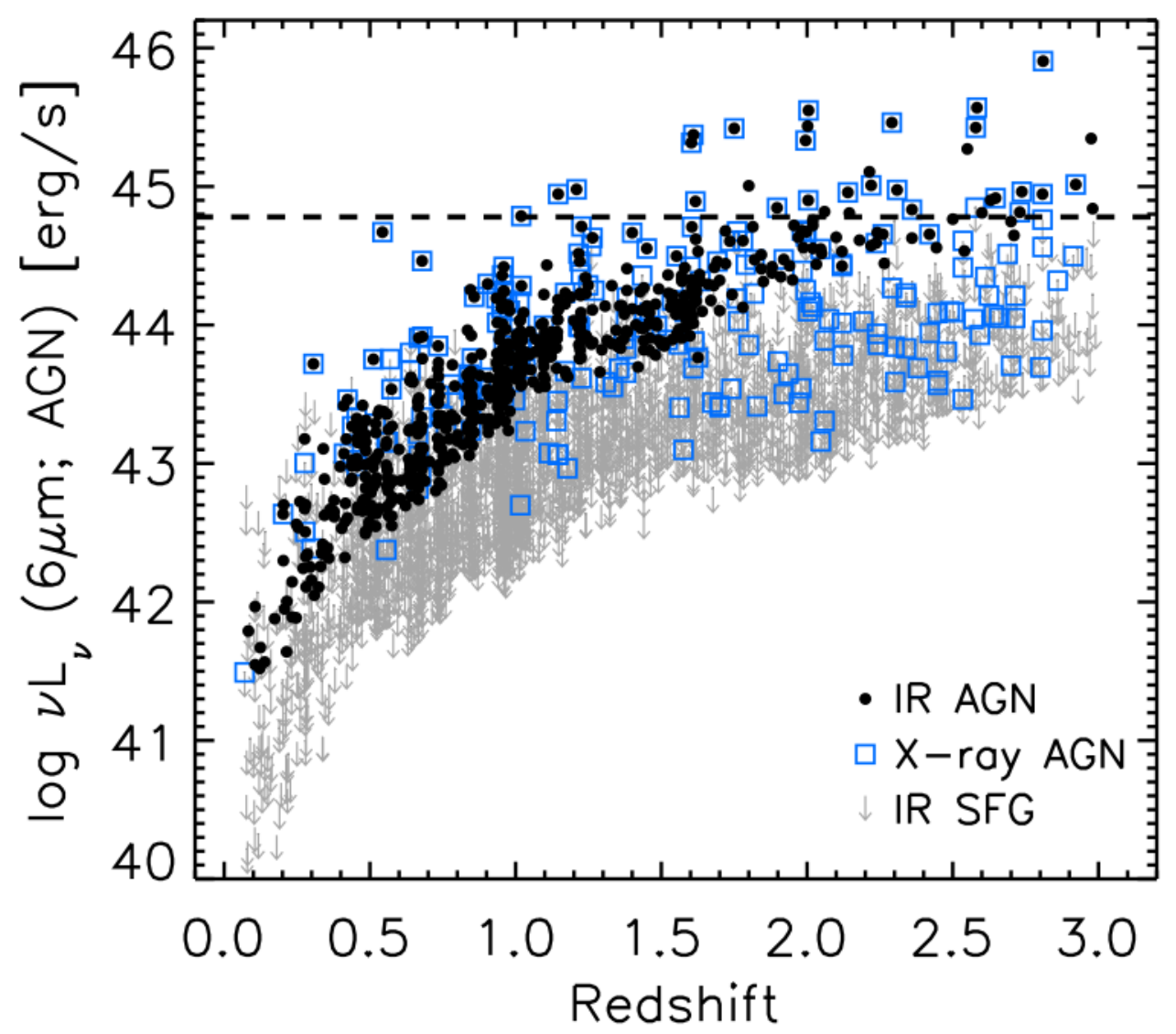}}
\caption{Rest-frame 6~$\mu$m AGN luminosity, calculated from the SED fits, vs. redshift for all of the 24~$\mu$m detected sources in the GH fields ($\sim$3200 sources with $z\le3$, out of the total 3690 24~$\mu$m-detected sources). The black circles represent the sources where we significantly identify an AGN component in the IR SED (Sect. \ref{sed}), while the blue open squares represent the X-ray detected AGN. For sources where an AGN component was not found in the best-fitting SEDs (``IR SFG''), the 6~$\mu$m AGN luminosity is plotted as an upper limit (grey arrows). The dashed line at $\log \nu L_{\rm 6 \mu m}=44.78$~\ergs\ marks our threshold for the IR quasar selection.}
\label{fig.l6z}
\end{figure}

After identifying the best-fitting model, we derived the luminosity of the AGN (i.e., corrected for the galaxy contribution) at rest-frame 6~$\mu$m ($L_{\rm 6 \mu m,\ AGN}$), the IR luminosity of the AGN ($L_{\rm AGN}$) calculated across the wavelength range $\lambda=$~8--1000~$\mu$m, and the IR luminosity of the SFG ($\lambda=$~8--1000~$\mu$m; hereafter $L_{\rm SF}$), as weighted means of the values obtained from the five best-fitting model solutions (or from all the solutions satisfying condition (1) for the AGN $+$ SFG model). 
Where the AGN component was not significantly identified an upper limit to the AGN luminosity at 6~$\mu$m was set as 50\% of the total 6~$\mu$m emission (Figure \ref{fig.l6z}), i.e., the limit for the SFG component becoming dominant at 6~$\mu$m. 
For the cases where the FIR data (100, 160 and 250 $\mu$m) are all upper limits and our SED fitting approach did not provide a reliable measurement of the SF emission, we estimated an upper limit for  $L_{\rm SF}$ by increasing the normalisation of the SF templates until they reached any of the FIR upper limits, without exceeding any of the MIR data point measurements (including errors). An IR luminosity was calculated from each of the five templates and the highest value was then taken as the $L_{\rm SF}$ upper limit. 

\begin{table*}
\caption{IR quasar sample}\label{tab1}
\begin{small}
\begin{center}
\setlength{\tabcolsep}{0.15cm}
\begin{tabular}{c c c c c c c c c c c c c c}
\hline
\rule[-1.5mm]{0pt}{3ex}No &Coordinates & XID & Redshift & $L_{\rm X}$ & $\nu L_{\rm 6 \mu m,AGN}$ & $f_{\rm  AGN}$ & ${L_{\rm AGN}}$& ${L_{\rm bol}}$& $L_{\rm SF}$ & $SFR$ & ${ L_{\rm X,SF}}$& ${ f_{\rm X,SF}}$ & Opt. class\\
\rule[-1.5mm]{0pt}{3ex}(1) & (2) & (3) & (4) & (5) & (6) & (7) & (8) & (9) & (10) & (11) &{(12)} &{(13)} & (14)\\
\hline
\rule[-1.5mm]{0pt}{3ex}   1 & 12:36:00.16\ $+$62:10:47.4       & --   & 2.002 s &$<$42.30 & 45.44 & {80\%} & {46.01} &{46.31} &  46.21 &   723    & {41.61}   &{$>$20\%}&  \\
\rule[-1.5mm]{0pt}{3ex}   2 & 12:36:22.94\ $+$62:15:26.5       & 137N & 2.583 s &   44.80 & 45.57 & {100\%}& {46.09} &{46.39} & $<$45.65 & $<$197 & {$<$41.19}&{0\%} & BL\\
\rule[-1.5mm]{0pt}{3ex}   3 & 12:36:32.55\ $+$62:07:59.8       & 171N & 1.995 s &   42.69 & 45.33 & {94\%} & {45.85} &{46.15} & $<$45.22 & $<$73  & {$<$40.87}&{$<$2\%}  & HEG\\
\rule[-1.5mm]{0pt}{3ex}   4 & 12:36:35.60\ $+$62:14:24.0       & 190N & 2.005 s &   42.90 & 45.55 & {83\%} & {46.13} &{46.43} &    46.29 &   858  & {41.66}&{6\%}  & HEG\\
\rule[-1.5mm]{0pt}{3ex}   5 & 12:36:42.19\ $+$62:17:11.2       & 223N & 2.730 s &   43.93 & 44.81 & {76\%} & {45.53} &{45.83} &    45.70 &   221  & {41.23}& {0\%} & \\
\rule[-1.5mm]{0pt}{3ex}   6 & 12:36:46.72\ $+$62:14:45.9$^{a}$ & 243N & 2.004 s &   42.05 & 44.90 & {74\%} & {45.44} &{45.74} &   45.84 &   306   & {41.33 }& {19\%}& \\ 
\rule[-1.5mm]{0pt}{3ex}   7 & 12:36:49.65\ $+$62:07:38.1       & 259N & 1.610 s &   44.04 & 45.37 & {94\%} & {45.92} & {46.22} &    45.51 &   145  & {41.09}&{0\%} & HEG\\
\rule[-1.5mm]{0pt}{3ex}   8 & 12:36:55.82\ $+$62:12:01.2       & 287N & 2.737 s &   43.06 & 44.96 & {69\%} & {45.66} & {45.96} &    45.94 &   390  & {41.41}& {2\%}& \\
\rule[-1.5mm]{0pt}{3ex}   9 & 12:36:56.47\ $+$62:19:37.6       & --   & 1.80  p &$<$42.33 & 45.01 & {73\%} & {45.54} & {45.84} &    45.95 &   396  & {41.41}& {$>$19\%}& \\
\rule[-1.5mm]{0pt}{3ex}  10 & 12:36:57.93\ $+$62:21:28.7       & 299N & 2.36  p &   44.12 & 44.83 & {74\%} & {45.16} & {45.46} &    45.88 &   332  & {41.36}&{0\%} & \\
\rule[-1.5mm]{0pt}{3ex}  11 & 12:36:59.32\ $+$62:18:32.4       & 307N & 2.14  p &   42.74 & 44.96 & {83\%} & {45.68} & {45.98} &    45.64 &   195  & {41.18}&{3\%} & \\
\rule[-1.5mm]{0pt}{3ex}  12 & 12:37:04.34\ $+$62:14:46.2       & --   & 2.214 s &$<$42.09 & 45.11 & {91\%} & {45.85} & {46.15} &    45.41 &   114  & {41.01}&{$>$4\%}  & HEG\\
\rule[-1.5mm]{0pt}{3ex}  13 & 12:37:06.87\ $+$62:17:02.1       & 344N & 1.019 s &   44.14 & 44.79 & {90\%} & {44.92} & {45.22} &    45.17 &   65   & {40.83}&{0\%}  & BL \\
\rule[-1.5mm]{0pt}{3ex}  14 & 12:37:16.67\ $+$62:17:33.3       & 390N & 1.146 s &   43.70 & 44.95 & {98\%} & {45.47} & {45.77} &    44.58 &   17   & {40.39}&{0\%} & SF\\
\rule[-1.5mm]{0pt}{3ex}  15 & 12:37:19.87\ $+$62:09:55.2       & 398N & 2.647 s &   43.77 & 44.92 & {85\%} & {45.45} & {45.75} &    45.54 &   155  & {41.11}&{0\%} & SF\\
\rule[-1.5mm]{0pt}{3ex}  16 & 12:37:26.50\ $+$62:20:26.6       & 423N & 1.750 s &   42.74 & 45.42 & {89\%} & {46.12} & {46.42} &    45.84 &   303  & {41.33}&{4\%} & \\
\rule[-1.5mm]{0pt}{3ex}  17 & 12:37:39.50\ $+$62:15:58.6       & --   & 2.98  p &$<$42.76 & 44.84 & {66\%} & {45.57} & {45.87} &    45.97 &   414  & {41.43}&{$>$5\%} & \\
\rule[-1.5mm]{0pt}{3ex}  18 & 12:37:42.53\ $+$62:18:11.8       & 459N & 2.309 s &   44.70 & 44.98 & {93\%} & {45.25} & {45.55} & $<$45.30 & $<$88  & {$<$40.93}&{0\%} & BL\\
\rule[-1.5mm]{0pt}{3ex}  19 & 12:37:57.28\ $+$62:16:27.4       & 478N & 2.922 s &   44.41 & 45.01 & {66\%} & {45.48} & {45.78} &    46.01 &   458  & {41.46}&{0\%} & BL\\
\rule[-1.5mm]{0pt}{4.5ex}20 & 03:32:09.45\ $-$27:48:06.7       & 149S & 2.810 s &   43.96 & 45.91 & {100\%} & {46.22} & {46.52} & $<$45.80 & $<$280& {$<$41.30}&{0\%} & BL\\ 
\rule[-1.5mm]{0pt}{3ex}  21 & 03:32:11.78\ $-$27:46:28.2       & 176S & 2.81  p &   43.81 & 44.95 & {92\%} & {45.12} & {45.42} & $<$45.39 & $<$109 & {$<$41.00}& {0\%}& \\
\rule[-1.5mm]{0pt}{3ex}  22 & 03:32:20.05\ $-$27:44:47.2       & 278S & 1.897 s &   42.43 & 44.85 & {73\%} & {45.57} & {45.87} &    45.82 &   291  & {41.31}&{8\%} & HEG\\
\rule[-1.5mm]{0pt}{3ex}  23 & 03:32:23.44\ $-$27:42:55.2       & 320S & 2.145 s &   42.24 & 44.81 & {80\%} & {45.50} & {45.80} &    45.59 &   174  & {41.15}&{8\%} & \\
\rule[-1.5mm]{0pt}{3ex}  24 & 03:32:23.72\ $-$27:44:11.8       & --   & 2.060 s &$<$42.00 & 44.82 & {83\%} & {45.38} & {45.68} &    45.34 &   97   & {40.96}&{$>$9\%} & \\
\rule[-1.5mm]{0pt}{3ex}  25 & 03:32:24.49\ $-$27:50:45.8       & --   & 2.630 s &$<$42.06 & 44.90 & {88\%} & {45.63} & {45.93} & $<$45.39 & $<$108 & {$<$40.99}& {9\%}& \\
\rule[-1.5mm]{0pt}{3ex}  26 & 03:32:25.14\ $-$27:42:19.2       & 344S & 1.617 s &   43.77 & 44.89 & {66\%} & {45.44} & {45.74} &    46.00 &   438  & {41.44}& {0\%}& BL\\
\rule[-1.5mm]{0pt}{3ex}  27 & 03:32:25.69\ $-$27:43:05.7       & 351S & 2.291 s &   43.79 & 45.46 & {100\%} & {45.74} & {46.04} & $<$45.37 & $<$103& {$<$40.98}&{0\%} & HEG\\
\rule[-1.5mm]{0pt}{3ex}  28 & 03:32:28.82\ $-$27:48:29.7       & --   & 2.550 s &$<$41.95 & 45.27 & {78\%} & {45.97} & {46.27} &    45.93 &	375    & {41.39}& {$>$28\%}& \\
\rule[-1.5mm]{0pt}{3ex}  29 & 03:32:31.47\ $-$27:46:23.2       & 435S & 2.220 s &   42.44 & 45.01 & {75\%} & {45.30} & {45.60} &    45.95 &   397  & {41.41}&{9\%} & HEG\\
\rule[-1.5mm]{0pt}{3ex}  30 & 03:32:35.72\ $-$27:49:16.0       & 490S & 2.579 s &   42.53 & 45.43 & {78\%} & {46.08} & {46.38} &    46.29 &   866  & {41.66}& {14\%}& HEG\\
\rule[-1.5mm]{0pt}{3ex}  31 & 03:32:37.76\ $-$27:52:12.2       & 518S & 1.603 s &   44.17 & 45.32 & {94\%} & {45.67} & {45.97} & $<$45.61 & $<$180 & {$<$41.16}& {0\%}& HEG\\
\rule[-1.5mm]{0pt}{3ex}  32 & 03:32:37.79\ $-$27:42:32.8       & --   & 2.975 s &$<$42.57 & 45.35 & {96\%} & {46.07} & {46.37} & $<$45.78 & $<$269 & {$<$41.29}& {5\%}& \\
\rule[-1.5mm]{0pt}{3ex}  33 & 03:32:49.58\ $-$27:47:14.9       & --   & 2.600 s &$<$42.33 & 44.81 & {94\%} & {45.54} & {45.84} & $<$45.26 & $<$81  & {$<$40.90}&{4\%} & \\
\hline
\hline
\end{tabular}
\end{center}
\end{small}
NOTES: (1) Source number; (2) MIR 3.6~$\mu$m source position, RA and Declination; (3) X-ray source identification number from \citet{alexander2003} in CDF-N and from \citet{xue2011} in CDF-S; the suffix ``N'' and ``S'' refer to North and South, respectively; (4) Redshift; ``s'' and ``p'' indicate the spectroscopic and photometric redshifts, respectively (Sect. \ref{ir}); (5) Logarithm of the X-ray luminosity at rest-frame 2-10~keV{, extrapolated from the measured 0.5-2 keV flux (or upper limit),} in units of~\ergs; (6) Logarithm of the rest-frame 6~$\mu$m luminosity of the AGN measured from the best-fitting SEDs in units of~\ergs; { (7) Contribution from the AGN emission to the total (AGN $+$ SF) emission at 6~$\mu$m; (8) Logarithm of the AGN IR luminosity (8--1000~$\mu$m) in units of~\ergs; (9) Logarithm of the AGN bolometric luminosity, derived from $L_{\rm AGN}$, in units of \ergs; } (10) Logarithm of the IR luminosity (8--1000~$\mu$m) of the star-formation component in units of~\ergs; (11) Star-formation rate in units of $\rm M_{\odot}~yr^{-1}$; {(12) Logarithm of the rest-frame 2-10~keV luminosity produced by SF, estimated assuming the \citet{lehmer2010} relation between SFR and hard X-ray luminosity; (13) Estimated contribution from SF to the rest-frame 2-10~keV luminosity (uncorrected for absorption);} (14) Optical spectroscopic classification from \citet{steidel2002}, \citet{barger2003} and \citet{trouille2008} for the sources in GH-N and from \citet{szokoly2004} and \citet{silverman2010} in GH-S: broad-line AGN (BL), high-excitation emission line galaxies (HEG); star-forming galaxies (SF). 
$^{a}$ For this source we report the X-ray coordinates as the 3.6~$\mu$m coordinates are centred on a low-redshift galaxy ($z=0.556$) lying $\approx1.8''$ from the X-ray source. Careful inspection of the multi-wavelength images has proved that although the \spz-IRAC bands are dominated by the low-$z$ galaxy emission, the emission at $\lambda\ge8\ \mu$m is centred at the X-ray source position rather than the 3.6~$\mu$m position of the low-$z$ galaxy. We therefore expect the contamination from the low-z galaxy to be very small at the wavelengths used in our SED analyses of this IR quasar (see Sect. \ref{sed}). 
\end{table*}

\subsection{IR quasar sample}\label{irquasar}

From the results obtained from our SED decomposition, we selected a sample of luminous AGN with $\nu L_{\rm 6\mu m,~AGN}\ge6\times10^{44}$~\ergs ($\ge44.78$ in the logarithmic scale; Fig. \ref{fig.l6z}), defined hereafter as our ``IR quasar'' sample. Considering the intrinsic relation between the 6~$\mu$m luminosity and the X-ray luminosity observed for unobscured AGN (e.g., \citealt{lutz2004,fiore2009,gandi2009,mateos2015}), our 6~$\mu$m luminosity cut corresponds to an intrinsic X-ray luminosity $L_{\rm 2-10~keV}\approx 2.5\times10^{44}$~\ergs, i.e., clearly in the quasar regime even allowing for the $\approx$0.4~dex scatter in the relation. We also restricted our sample to the redshift range $z=$~1--3 as we aim to investigate the population of luminous quasars at the peak of activity at $z\approx2$ (\citealt{madau1996, hopkins2006a, ueda2014, brandt2015}). 
Our sample consists of 33 sources, 19 in the GH-N and 14 in the GH-S fields (see Table \ref{tab1}). { The bolometric luminosities ($L_{\rm bol}$) of these quasars, estimated as $L_{\rm bol}~{\approx}~2\times L_{\rm AGN}$ { (assuming a bolometric correction of a factor of ${\sim8}$ at 6~$\mu$m, from \citealt{richards2006}, and considering the median ratio of ${\sim0.3}$ between ${\nu L_{\rm 6\mu m,~AGN}}$ and ${L_{\rm AGN}}$ estimated for our quasars)}, span the range $\log\ L_{\rm bol}=45.4-46.5$~\ergs\ (see Table \ref{tab1})}. For all of these sources we find that the AGN component { largely} dominates over the SF component at MIR wavelengths, with the AGN contributing $>${66\% to the total emission} at 6~$\mu$m  { ($f_{\rm AGN}=66-100$\%; where $f_{\rm AGN}$ is defined as the fraction of AGN luminosity over the total at rest-frame 6~$\mu$m; see Table \ref{tab1})}. The best-fitting SEDs are reported in Appendix \ref{ap1} (Fig. \ref{fig.sed}, \ref{fig.sed1}). We note that for 13 out of the 33 sources ($\approx$40\%) \spz-IRS low-resolution spectroscopy is available ($\lambda\approx$~3-20~$\mu$m; \citealt{pope2008b, murphy2009, kirkpatrick2012}). Although we do not use these spectra to constrain the source SEDs, we find very good agreement between our SED best-fitting solutions and the \spz-IRS spectra, which are shown in Fig. \ref{fig.sed}, \ref{fig.sed1} (Appendix \ref{ap1}), confirming the reliability of our SED fitting approach.

To explore the completeness of the sample we can take into account that when the AGN is not dominant in the MIR band, i.e., its contribution to the total emission is $<50$\%, our SED fitting procedure might start failing in identifying the presence of an AGN. We therefore looked for all the sources identified as SFGs from our SED fitting procedure (i.e., with no significant AGN component) with AGN luminosity upper limits above our IR quasar selection threshold of $\log\ \nu L_{\rm 6 \mu m,\ AGN}\ge44.78$~\ergs. There is only one SFG satisfying this condition, having $\log\ \nu L_{\rm 6 \mu m,\ AGN}<44.85$~\ergs. This source is classified as an AGN from its X-ray emission ($\log L_{\rm X}\approx43.2$~\ergs). Given its modest X-ray luminosity and since the $L_{\rm 6 \mu m,\ AGN}$ upper limit is very close to our IR quasar luminosity threshold (only a factor $\sim$1.2 higher), the AGN within this SFG might still not fulfil our selection criteria, i.e., its intrinsic power could be below our luminosity cut. For this reason we do not include this source in our sample, keeping in mind that our selection might not be 100\% complete in identifying IR quasars when the SF luminosity is very high.  

We note, however, that our IR quasar selection through the SED analysis is more reliable than the typical MIR colour-colour selections used to identify AGN (e.g., \citealt{stern2005,lacy2007,donley2012}), as the SED analysis accounts for the source redshifts and corrects for any contamination from SF to the MIR emission, contrary to any of the colour-colour selection techniques. Indeed, despite our sources being the most luminous quasars at MIR wavelengths within the GH fields, the \citet{stern2005} AGN wedge would miss $\approx$20\% of them, as well as the majority of the less luminous AGN that we identified from the SED analysis (see Fig. \ref{fig.l6z}), probably due to the contamination from SF and/or starlight to the emission in the MIR bands used in the colour-colour plot (typically \spz-IRAC bands). Also the \citet{donley2012} AGN selection criteria would miss several of our IR quasars, selecting only $\sim$73\% of them. On the other hand, the \citet{lacy2007} AGN wedge would select $\sim$97\% (all but one) of the IR quasars in our sample; however, a large fraction of the sources lying within the wedge are not identified as AGN according to our SED analysis, nor in the X-ray band, as SFGs at low and high redshifts can enter the \citet{lacy2007} AGN locus, introducing significant contamination from non-AGN (e.g., \citealt{donley2012}).

Of our final sample of 33 luminous IR quasars, 24 ($\sim$73\%) are detected in the X-ray band (Sect. \ref{xdata}). The X-ray identification number of these sources (XID) and their rest-frame 2-10~keV luminosity ($L_{\rm X}$) are reported in Table \ref{tab1}. These luminosities are extrapolated from the observed 0.5-2~keV fluxes reported in the \citet{alexander2003} and \citet{xue2011} catalogues for the GH-N and GH-S sources, respectively, using a photon index $\Gamma=1.8$ and appropriate k-correction. { This simple extrapolation is performed to have a first idea of the discrepancy between the observed X-ray and the MIR emission of our quasars. We caution, however, that this X-ray luminosity calculation does not account for the possible contribution from star formation to the observed X-ray flux { (see Table \ref{tab1} and section \ref{sfr})}. Accurate estimates of the observed and intrinsic 2-10~keV luminosity for the X-ray detected sources, derived from spectral analyses, are obtained in Section \ref{xray} and reported in Table \ref{tab2}; we use these luminosities for further analyses throughout the paper.}

\section{Analyses and results}\label{results}

In this Section we present the X-ray spectral analysis performed on all of the X-ray detected IR quasars (24 sources) to constrain the absorbing column density (\nh) and therefore derive the intrinsic X-ray luminosity. We then estimate the obscured quasar (\nh$>10^{22}$~cm$^{-2}$) and CT quasar (\nh$>1.5\times10^{24}$~cm$^{-2}$) fractions from the X-ray spectral fitting results and from a comparison between the observed (i.e., not corrected for absorption) 2--10~keV luminosity and the $L_{\rm 6 \mu m,\ AGN}$ (rest-frame), used as a proxy of the intrinsic AGN power. We then investigate in detail the host galaxy properties through the SFRs estimated from the { $L_{\rm SF}$} derived from the SED fitting analysis, in relation to the AGN intrinsic luminosity and X-ray obscuration. We finally explore the impact of galaxy mergers/interactions in quasar hosts by visually classifying the host galaxy morphology disturbance of our IR quasars using high-resolution HST images. 

\subsection{X-ray spectroscopy}\label{xray}

\begin{table*}
\caption{X-ray spectral results for all of the X-ray detected sources}
\begin{center}
\begin{tabular}{c c c c c c c c c c}
\hline
\rule[-1.5mm]{0pt}{3ex}No & XID & Net Counts & $\Gamma_{\rm mo1}$ & $N_{\rm H,\ mo1}$ & $\Gamma_{\rm mo2}$ & $N_{\rm H,\ mo2}$ & $f_{\rm scatt}$ (\%)  &log $L_{\rm 2-10~keV}$ & log $L_{\rm 2-10~keV, intr}$ \\
\rule[-1.5mm]{0pt}{3ex}(1) & (2) & (3) & (4) & (5) & (6) & (7) & (8) & (9) & (10)\\
\hline
\rule[-1.5mm]{0pt}{3ex}2  & 137N & 3587 & 1.77$_{-0.05}^{+0.05}$ & $<$0.8		         & --         & --	   & --  & 44.72 & 44.72 \\ 
\rule[-1.5mm]{0pt}{3ex}3  & 171N &  128 & 0.28$_{-0.46}^{+0.74}$ & $<$15.3  	         & 1.80       & 74.3$_{-32.8}^{+65.3}$ & $7.4\pm1.4$  & 42.94 & 43.79 \\ 
\rule[-1.5mm]{0pt}{3ex}4  & 190N &  159 & 1.25$_{-0.74}^{+0.76}$ & 20.9$_{-15.4}^{+18.6}$& 1.82$_{-0.67}^{+0.65}$ & 49.7$_{-21.5}^{+25.3}$& $7.9\pm6.1$ & 43.13 & 43.82 \\ 
\rule[-1.5mm]{0pt}{3ex}5  & 223N &  421 & 1.81$_{-0.30}^{+0.33}$ & 9.1$_{-5.0}^{+6.1}$   & 1.91$_{-0.34}^{+0.41}$ & 10.8$_{-4.6}^{+17.1}$ & $7.4\pm11.6$ & 43.87 & 44.15 \\ 
\rule[-1.5mm]{0pt}{3ex}6  & 243N &   12 & 1.86$_{-0.87}^{+1.29}$ & --                    & 1.80       & $<$8.2	 & -- & 42.06 & 42.11 \\ 
\rule[-1.5mm]{0pt}{3ex}7  & 259N & 2952 & 1.86$_{-0.12}^{+0.13}$ & 8.2$_{-1.1}^{+1.2}$   & 1.93$_{-0.13}^{+0.14}$ & 10.6$_{-1.8}^{+1.9}$ & $4.2\pm1.3$ & 44.22 & 44.49 \\ 
\rule[-1.5mm]{0pt}{3ex}8  & 287N &   56 & 1.91$_{-0.50}^{+0.41}$ & $<$2.6		         & --         & --	   & --  & 43.00 & 43.00 \\ 
\rule[-1.5mm]{0pt}{3ex}10 & 299N &  795 & 1.66$_{-0.12}^{+0.21}$ & $<$1.5		         & --         & --	   & --  & 44.05 & 44.05 \\ 
\rule[-1.5mm]{0pt}{3ex}11 & 307N &   89 & 1.29$_{-0.83}^{+0.83}$ & 13.8$_{-12.8}^{+17.6}$& 1.71$_{-1.29}^{+1.39}$ & 39.2$_{-36.1}^{+54.4}$ & $7.2\pm6.5$ & 42.92 & 43.54 \\ 
\rule[-1.5mm]{0pt}{3ex}13 & 344N & 7007 & 2.05$_{-0.04}^{+0.04}$ & $<$0.8  	         & --         & --	     & -- & 44.03 & 44.03 \\ 
\rule[-1.5mm]{0pt}{3ex}14 & 390N & 2780 & 1.48$_{-0.10}^{+0.10}$ & 1.5$_{-0.4}^{+0.4}$ & 1.65$_{-0.14}^{+0.15}$ & 3.9$_{-1.5}^{+1.7}$ & $26.5\pm6.5$ & 43.89 & 44.00 \\ 
\rule[-1.5mm]{0pt}{3ex}15 & 398N &  414 & 1.76$_{-0.32}^{+0.35}$ & 14.5$_{-6.3}^{+7.7}$& 1.95$_{-0.48}^{+0.59}$ & 29.9$_{-17.6}^{+28.9}$ & $13.4\pm10.9$ & 43.75 & 44.12 \\ 
\rule[-1.5mm]{0pt}{3ex}16 & 423N &  114 & 0.54$_{-0.41}^{+0.56}$ & $<$5.6		      & 1.80      & 137.3$_{-105.1}^{+238.7}$ & $10.9\pm1.9$ & 42.92 & 43.84 \\ 
\rule[-1.5mm]{0pt}{3ex}18 & 459N & 3301 & 1.76$_{-0.06}^{+0.08}$ & $<$0.4		         & --         & --	   & --  & 44.62 & 44.62 \\ 
\rule[-1.5mm]{0pt}{3ex}19 & 478N &  890 & 2.05$_{-0.13}^{+0.20}$ & $<$1.6		         & --         & --	   & --  & 44.35 & 44.35 \\
\rule[-1.5mm]{0pt}{4.5ex}20& 149S& 1286 & 1.69$_{-0.18}^{+0.18}$ & 20.6$_{-4.9}^{+5.1}$& --         & --	  & --   & 43.91 & 44.37 \\ 
\rule[-1.5mm]{0pt}{3ex}21 & 176S &  677 & 1.93$_{-0.25}^{+0.27}$ & 13.2$_{-5.1}^{+6.0}$& 2.40$_{-0.48}^{+0.52}$  & 41.8$_{-24.6}^{+23.9}$ & $15.1\pm7.9$ & 43.70 & 44.07 \\
\rule[-1.5mm]{0pt}{3ex}22 & 278S &  165 & 1.35$_{-1.03}^{+1.01}$ & 68.7$_{-38.4}^{+44.9}$& 1.75$_{-1.21}^{+1.07}$& 89.8$_{-50.6}^{+47.4}$ & $0.8\pm0.5$ & 42.89 & 44.01 \\ 
\rule[-1.5mm]{0pt}{3ex}23 & 320S &   48 &$-$0.73$_{-0.97}^{+0.59}$ & -- 	           & 1.80       & 234.8$_{-104.2}^{+176.0}$ & $1.5\pm0.2$ & 41.95 & 43.65 \\
\rule[-1.5mm]{0pt}{3ex}26 & 344S & 1680 & 2.10$_{-0.14}^{+0.16}$ & 0.7$_{-0.5}^{+0.6}$ & --         & --	   & --  & 43.69 & 43.73 \\ 
\rule[-1.5mm]{0pt}{3ex}27 & 351S & 1378 & 1.70$_{-0.17}^{+0.18}$ & 13.7$_{-2.7}^{+3.1}$& 1.76$_{-0.20}^{+0.24}$  & 16.4$_{-4.2}^{+7.6}$ & $3.8\pm4.9$ & 43.82 & 44.15 \\ 
\rule[-1.5mm]{0pt}{3ex}29 & 435S &   59 & 1.11$_{-0.44}^{+0.46}$ & $<$0.8		       & 1.89$_{-0.61}^{+0.67}$ & 1100.2$_{-695.8}^{+1235.4}$ & $2.8\pm2.9$ & 42.46 & 44.03 \\ 
\rule[-1.5mm]{0pt}{3ex}30 & 490S &   82 & 0.37$_{-0.76}^{+1.76}$ & $<$76.0             & 1.80       & 2591.4$_{-1464.5}^{+2037.2}$ & $0.5\pm0.1$ & 42.55 & 44.86 \\ 
\rule[-1.5mm]{0pt}{3ex}31 & 518S & 4961 & 2.17$_{-0.08}^{+0.08}$ & 2.1$_{-0.3}^{+0.4}$ & --         & --	  & --   & 44.17 & 44.27 \\ 
\hline
\hline 
\end{tabular}
\end{center}
NOTES: (1) Source number as in Table \ref{tab1} ; (2) XID as in Table \ref{tab1}; (3) Net counts in the \ch\ spectra in the observed 0.5-8~keV energy range; (4) Photon index, which was left free to vary in our spectral fits (model 1); (5) Intrinsic hydrogen column density in units of $10^{22}$~cm$^{-2}$; (6) Photon index obtained from model 2; $\Gamma=1.80$ has been fixed where the spectral fits could not reliably constrain the spectral slopes and \nh\ simultaneously (Sect. \ref{mo1}); (7) Hydrogen column density obtained from model 2, in units of $10^{22}$~cm$^{-2}$; (8) Fraction of the intrinsic power-law emission that is scattered at soft energies ($E<2$~keV); (9) Logarithm of the rest-frame 2-10~keV luminosity, in units of~\ergs; (10) Logarithm of the intrinsic, i.e. corrected for absorption, 2-10~keV luminosity (rest frame), in units of~\ergs.
\label{tab2}
\end{table*}%

For all of the 24 IR quasars with an X-ray detection in the CDF-N and CDF-S catalogues (\citealt{alexander2003,xue2011}) we performed X-ray spectral analyses to investigate the high energy properties of these sources. For the X-ray undetected sources we estimated 3$\sigma$ flux upper limits through aperture photometry on the \ch\ images, assuming a spectral slope of $\Gamma=1.4$ for the count rate to flux conversion. The 2-10~keV luminosity upper limits derived from these fluxes are reported in Table \ref{tab1}.

The spectra were extracted using the \emph{ACIS Extract} (AE) software package \citep{broos2010,broos2012} and grouped with a minimum of 20 counts per energy bin in order to adopt $\chi^2$ fitting statistics. The spectra with limited counting statistics ($<200$ net counts at 0.5-8.0~keV) were binned with a minimum of one count per bin and the Cash statistic \citep{cash1979} was adopted for the spectral analysis with XSPEC (v. 12.8.1g). 
\begin{figure}
\centerline{
\includegraphics[scale=0.63]{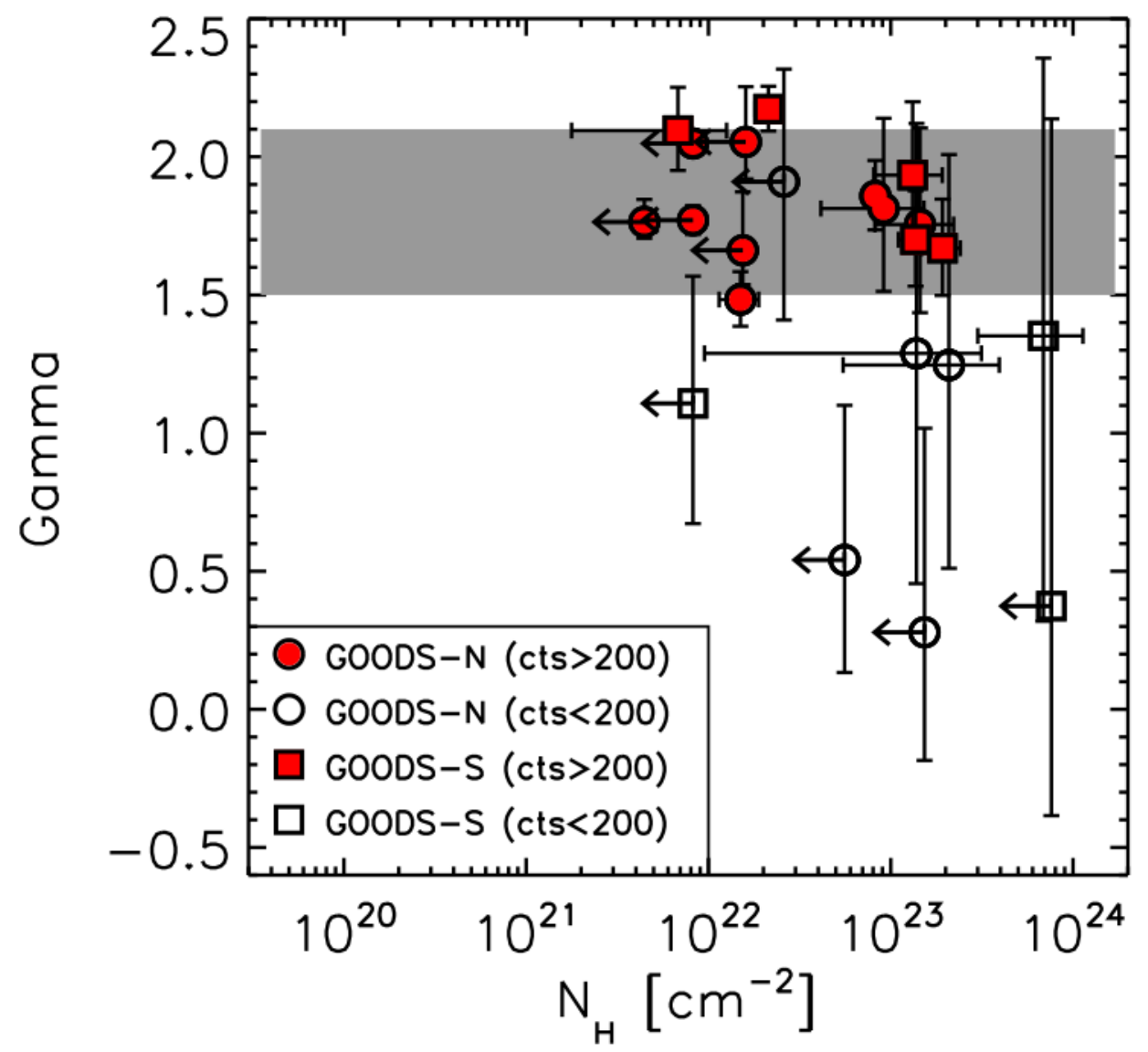}}
\caption{Photon index $\Gamma$ vs intrinsic hydrogen column density ($N_{\rm H}$) for the X-ray detected IR quasars in our sample, obtained using a simple absorbed power-law model (model 1); the sources XID 243N and XID 320S are not included in the plot because the $\Gamma$ and \nh\ parameters could not be simultaneously constrained from the spectral fit due to the low numbers of counts ($<50$ cts; see Sect. \ref{xray}). Sources in GH-N and GH-S are represented by circles and squares, respectively. Red filled symbols indicate sources with more than 200 net counts in the $E=$~0.5--8~keV spectra; open symbols are sources with $<$200 net counts (i.e., with poor counting statistics). The grey shaded region represents the intrinsic photon indices typically found for AGN ($\Gamma=1.8\pm0.3$; e.g., \citealt{nandra1994,mainieri2002,mateos2005b,tozzi2006}). The quasars with low numbers of counts have typically flat photon indices ($\Gamma<1.5$), suggesting the presence of additional spectral components, such as soft scattered emission and/or reflection (Sect. \ref{mo1}).}
\label{fig.nhgamma}
\end{figure}

\subsubsection{Basic spectral model fitting}\label{mo1}
We initially used a simple absorbed power-law model to characterise the source spectra (model 1), in addition to Galactic absorption, which was fixed to the average values of $N_{\rm H, Gal}=10^{20}$~cm$^{-2}$ in GH-N and $N_{\rm H, Gal}=9.0\times10^{19}$~cm$^{-2}$ in GH-S (\citealt{dickey1990}). We left the photon index $\Gamma$ and the intrinsic hydrogen column density (\nh) as free parameters. In Figure \ref{fig.nhgamma} we show the $\Gamma$ versus \nh\ obtained from these fits. For two quasars (XID 243N  and XID 320S; see Table \ref{tab1}) with very low numbers of counts ($<50$ net counts, 0.5--8.0~keV), we only fit a simple power-law model with no intrinsic absorption as no useful constraints could be obtained simultaneously on $\Gamma$ and \nh\ from model 1. These two sources are not plotted in Fig. \ref{fig.nhgamma}. The results of the X-ray spectral analysis (including errors at 90\% confidence level) are reported in Table \ref{tab2}. We note that 14 of the 24 analysed sources ($\sim$58\%) have good counting statistics (net counts $>200$; see Table \ref{tab2}). For these sources we obtained good spectral constraints with our simple model, with spectral slopes consistent with the intrinsic values typically found for AGN ($\Gamma=1.8\pm0.3$; e.g., \citealt{nandra1994,mainieri2002,mateos2005b,tozzi2006}). The sources with lower counting statistics (net counts $<200$) have typically flatter spectral indices, ranging between $\Gamma\approx$~0.3--1.4, with large errors. This indicates that the spectra of these sources are harder, and likely more obscured than those measured for the sources with higher numbers of counts; however the poor counting statistics do not allow us to  simultaneously place tight constraints on the spectral parameters, $\Gamma$ and \nh, as they are degenerate for low count spectra (e.g., \citealt{mateos2008}). The flat $\Gamma$ in these sources might also indicate that a simple absorbed power-law model does not properly characterise their X-ray emission and additional spectral components might be required (e.g., soft scattered component, or reflection). 
 
For the sources with a flat photon index ($\Gamma<1.5$) we therefore included an extra power-law component in the model to fit the emission at soft energies ($E\lesssim1$~keV; hereafter model 2). There are several processes that can cause emission at soft X-ray energies, such as thermal emission from star forming regions, radio jets, atomic processes in partially ionised gas, { partial covering by a neutral absorber, or} blurred ionised reflection from the inner parts of the accretion disc (e.g., {\citealt{turner1997}}; \citealt{gierlinski2004}); however, given the limited numbers of counts available in our spectra it is not possible to fit more complex models. We therefore assume a power-law model as an approximation of the soft X-ray emission, { with the photon index $\Gamma$ tied to that of the primary power law (e.g., \citealt{brightman2014, buchner2014, lanzuisi2015}) to limit the number of free parameters in the model.}\footnote{{ We also tested the spectral fits of model 2 with a soft power-law slope free to vary, but no useful constraints could be obtained on $\Gamma$ for { our quasars}, due to the limited counting statistics and the high redshift of the sources, which shifts most of the soft energy emission ($E\lesssim1-2$ keV, rest frame) out of the \ch\ sensitivity range.}} Fitting the spectra with model 2 constrained the power-law slopes to more typical values for AGN (see Table \ref{tab2}), although the uncertainties on the parameters are still large. Where reasonable constraints on $\Gamma$ could not be obtained, we fixed $\Gamma=1.8$ to obtain better constraints on the column densities. The \nh\ values resulting from the fitting of model 2 are typically higher than those obtained from model 1 and in some cases they are very high (\nh$>8\times10^{23}$~cm$^{-2}$) and possibly consistent with Compton-thick absorption. In two cases (XID 435S and XID 490S; see Sect. \ref{xsrc}) the column densities estimated by model 2 even exceed \nh$>10^{25}$~cm$^{-2}$. Although these \nh\ values are not reliable because the simple parameterisation of the spectra with an absorbed power law is not suitable to constrain column densities of \nh$>\rm few\times10^{23}$~cm$^{-2}$, they still indicate that these sources might indeed be heavily obscured, possibly CT. We investigate this hypothesis using more suitable models in Section \ref{ct}. In model 2 we also estimated the fraction of intrinsic emission scattered at soft energies ($f_{\rm scatt}$), calculated from the ratio between the normalisations of the scattered and intrinsic power-law components, which is on average only a few percent, typically $\lesssim$5\% (e.g., \citealt{turner1997,ueda2007}). However, in some cases we find higher scattered fractions (see Table \ref{tab2}), which might be due to the contribution from SF emission to the soft energy spectrum ($E<$~1-2~keV).

We also tested model 2 for sources showing significant X-ray absorption (\nh$>10^{22}$~cm$^{-2}$) from the fitting of model 1, but with steeper spectral slopes ($\Gamma>1.5$); the results are reported in Table \ref{tab2} for all of the sources where the normalisation of the soft scattered power law could be constrained. Although the column densities obtained from these fits are typically higher than those obtained from model 1, the scatter on the parameters ($\Gamma$ and \nh) is larger and in all cases the results from the two models are consistent. We therefore use the best-fitting solutions from model 1 to estimate the luminosity for these sources. 

From the best-fitting models (model 1 for the sources with $\Gamma_{\rm mo1}>1.5$, and model 2 for the remainders) we calculated the rest-frame 2-10~keV luminosity of the X-ray detected IR quasars ($L_{\rm 2-10~keV}$), as well as the unabsorbed 2-10~keV luminosity ($L_{\rm 2-10~keV, intr}$), i.e., corrected for the best estimate of the \nh\ (Table \ref{tab2}).

\begin{table}
\caption{Spectral fits of the heavily obscured quasars using PLCABS (\citealt{yaqoob1997}) and Torus (\citealt{brightman2011})}
\begin{center}
\setlength{\tabcolsep}{0.15cm}
\begin{tabular}{l l c c c c}
\hline
    & & \multicolumn{2}{c}{PLCABS} & \multicolumn{2}{c}{Torus$^a$} \\
\rule[-1.5mm]{0pt}{4ex}{ No} & XID & $\Gamma^b$ & \nh$^c$ & $\Gamma^b$ & \nh$^c$ \\
\hline
\rule[-1.5mm]{0pt}{3ex}{ 3} & 171N & 1.8                    & 79.1$^{+54.4}_{-39.4}$    & 1.8 & $>$332.5 \\
\rule[-1.5mm]{0pt}{3ex}{ 4} & 190N & 1.76$^{+0.70}_{-0.72}$ & 45.0$^{+24.9}_{-21.1}$    & 1.71$^{+0.71}_{-0.71}$ & 39.8$^{+27.6}_{-18.1}$ \\
\rule[-1.5mm]{0pt}{3ex}{11} & 307N & 1.8                & ${ 43.8^{+75.9}_{-26.0}}$& ${ 2.39^{+0.48}_{-0.39}}$& ${ >603.8^d}$ \\
\rule[-1.5mm]{0pt}{3ex}{16} & 423N & 1.8                    & 42.7$^{+231.1}_{-28.0}$   & 1.8 & 137.4$^{+195.9}_{-125.6}$ \\
\rule[-1.5mm]{0pt}{3ex}{20} & 149S$^d$ & 1.73$^{+0.16}_{-0.16}$ & 21.9$^{+4.0}_{-3.9}$      & 1.74$^{+0.14}_{-0.17}$ & 19.6$^{+3.2}_{-3.9}$ \\
\rule[-1.5mm]{0pt}{3ex}{22} & 278S & 1.8                    & 84.3$^{+18.8}_{-15.7}$    & 1.8 & 78.1$^{+18.4}_{-16.6}$ \\
\rule[-1.5mm]{0pt}{3ex}{23} & 320S & 1.8 				   & 181.2$^{+121.3}_{-67.2}$  & 1.8 & 159.8$_{-54.2}^{+153.7}$ \\ 
\rule[-1.5mm]{0pt}{3ex}{29} & 435S & 1.90$^{+0.70}_{-0.62}$ & 685.4$^{+885.0}_{-424.2}$ & 1.90$^{+0.84}_{-0.74}$ & 251.4$_{-135.9}^{+368.6}$\\
\rule[-1.5mm]{0pt}{3ex}{30} & 490S & 1.8 				   & 1013.4$^{+1962.7}_{-819.9}$& 1.8 & $>$429.5 \\ 
\hline\hline
\end{tabular}
\end{center}
{NOTES: $^a$ To fit the Torus model we fixed the opening angle to $\theta_{\rm tor}=60^{\circ}$ and the inclination angle to $\theta_{\rm inc}=80^{\circ}$, since these sources have typically poor counting statistics and these parameters cannot be constrained from the spectral fits. $^b$ Where sensible constraints could not be obtained simultaneously for the photon index $\Gamma$ and the column density \nh, we fixed $\Gamma=1.8$. $^c$ The column density \nh\ is expressed in units of $10^{22}$~cm$^{-2}$. $^d$ For this source we did not include any soft power-law component as it is not needed in the fit.}
\label{tab3}
\end{table}%

\subsubsection{Spectral models for heavily-obscured AGN}\label{ct}

From the spectral analysis using the simple models 1 and 2, for nine sources we obtained very large column density values (see Table \ref{tab2}) indicating that they are heavily obscured (defined here as: \nh$>2\times10^{23}$~cm$^{-2}$), or CT sources (\nh$>1.5\times10^{24}$~cm$^{-2}$). However, simple absorbed power-law models are not suitable to reliably constrain such large column densities, as they do not accurately account for the Compton scattering and reflection of the absorbed photons through an optically thick material. We therefore modelled the intrinsic emission of these sources using more appropriate models, such as: i) PLCABS (\citealt{yaqoob1997}), which simulates the attenuation of a power-law continuum by dense, cold matter, and ii) Torus (\citealt{brightman2011}), which is a Table model based on Monte Carlo simulations of X-ray radiative transfer, self-consistently taking into account Compton scattering and iron fluorescence emission. To represent the soft X-ray emission we used a power-law model, as in model 2. To fit the PLCABS model we left \nh\ and $\Gamma$ free to vary, with $\Gamma$ tied to that of the soft power-law component; where both parameters cannot be constrained due to low counting statistics, we fixed $\Gamma=1.8$. The maximum number of scatters in the model was set to 12 (see \citealt{yaqoob1997}), and all other parameters were fixed to their default values. For the Torus model we fixed the inclination angle to an edge-on view, $\theta_{\rm inc}=80^{\circ}$, and an opening angle\footnote{We note that we also tried the fits with $\theta_{\rm tor}=30^{\circ}$, however, due to the limited numbers of counts in the spectra it was not possible to determine which of the two geometries provided the best fit.} of $\theta_{\rm tor}=60^{\circ}$ (see \citealt{brightman2012}). The \nh\ and $\Gamma$ parameters were left free to vary, with $\Gamma$ tied to that of the soft scattering component, as for the PLCABS model, or fixed to $\Gamma=1.8$. The best-fitting parameters obtained with these two models are reported in Table \ref{tab3}. 

In general there is good agreement between the results obtained from the PLCABS and the Torus models. Although the parameters are not tightly constrained, as the errors are relatively large, all the model solutions resulting from our analyses indicate that these quasars are, in fact, heavily obscured and in some cases, as for XID 320S, XID 435S and XID 490S, even CT. We note that { three} more sources, XIDs 171N{, 307N} and 423N are consistent with being CT quasars within the uncertainties on the \nh\ parameter (Table \ref{tab3}).

\subsubsection{Notes on some individual sources}\label{xsrc}

Although most of the source spectra can be fitted with the simple models described in the previous Sections, and sensible results are obtained for their spectral parameters (see Tables \ref{tab2} and \ref{tab3}), some of the sources show some peculiar characteristics. Here we describe in more detail the most interesting sources amongst our X-ray detected IR quasar sample:
\begin{itemize}
\item[]\emph{XID 171N:} From the spectral fits using model 2 and PLCABS this $z=1.995$ source shows a very high level of absorption \nh$\approx8\times10^{23}$~cm$^{-2}$ ($\Gamma=1.8$), although not at CT levels. However, the residuals in the hard band ($E\gtrsim3$~keV, observed frame) suggest the presence of a reflection component and an emission line at $E\approx2$~keV (observed frame), typical of CT AGN spectra. Indeed, the spectral fit using the Torus model yields a column density lower limit which is well within the CT regime. \citet{georgantopoulos2009} have also classified this source as a CT AGN. The strong emission line is consistent with the redshifted iron K$\alpha$ line ($E=6.4$~keV) with an equivalent width of $EW=1.2^{+1.6}_{-1.1}\ \rm keV$; given the limited counting statistics of the spectrum the uncertainties on the line $EW$ are large, but still consistent with CT absorption. 
\item[]\emph{XID 243N:} The spectrum of this $z=2.004$ source has only 12 net counts ($0.5-8$~keV) therefore a more complex model than a simple power law cannot be fit to the data. The effective photon index measured from the spectrum is $\Gamma=1.86^{+1.29}_{-0.87}$. Fitting the spectrum with an absorbed power law with $\Gamma=1.8$ fixed to try to constrain the level of absorption results in an upper limit of \nh$<8\times10^{22}$~cm$^{-2}$, which is consistent with the source being unabsorbed. However, the X-ray luminosity of XID 243N is very low compared to the 6~$\mu$m luminosity measured from its IR SED (see Table \ref{tab1}) and the typical $L_{\rm X}-L_{\rm 6 \mu m }$ relations (e.g., \citealt{lutz2004,fiore2009,mateos2015}). The observed X-ray luminosity is consistent with a suppression of a factor of $\sim300$ with respect to the intrinsic luminosity measured in the MIR (see Table \ref{tab4}), and still more than a factor of $\sim$120 considering the intrinsic scatter of the $L_{\rm X}-L_{\rm 6 \mu m }$ relation ($\sim0.35$~dex; e.g., \citealt{mateos2015}). This suggests that XID 243N might be a heavily CT quasar (see Sect. \ref{obs} and Fig. \ref{fig.l6lx}) and that the X-ray emission seen by \ch\ is only the soft scattered component (typically a few percent of the intrinsic nuclear emission; e.g., \citealt{turner1997,ueda2007,young2007}, and references therein), or some emission due to SF in the host galaxy { (see Table \ref{tab1} and Section \ref{sfr})}. Another possibility is that this source is intrinsically X-ray weak, as it has been found, for instance, for broad absorption line (BAL) quasars (see e.g., \citealt{luo2014}). However, XID 243N has never been identified as a BAL quasar to date. Although the hypothesis of intrinsically weak X-ray emission cannot be ruled out for this quasar, we favour here the heavy-obscuration interpretation (see Sect. \ref{obs}).   
\item[]\emph{XID 287N:} Similar to XID 243N, this $z=2.737$ quasar has an apparently unobscured spectrum (with low numbers of counts) and an X-ray luminosity that is significantly smaller compared to the intrinsic AGN luminosity measured in the MIR band. Assuming a standard value for $L_{\rm X}/L_{\rm 6 \mu m }$ (e.g., \citealt{lutz2004,fiore2009,mateos2015}) we estimate that the observed X-ray luminosity is consistent with being suppressed by a factor of $\sim35$ (see Table \ref{tab4}). This suppression is consistent with the effect of CT absorption, but it is less extreme than that observed for XID 243N. Moreover, no other diagnostics have been found to confirm the intrinsic AGN luminosity measured in the MIR band, therefore the possibility of intrinsic X-ray weakness cannot be ruled out. 
\item[]\emph{XID 307N:} In the lightly binned spectrum (1 count/bin) of this quasar, clear residuals around the observed $\sim$2 keV indicate the presence of a strong iron emission line, consistent with being Fe~K$\alpha$ at the source redshift of $z=2.14$. We therefore added a Gaussian component to the fit, with $E=6.4$~keV and line width fixed at $\sigma=50$~eV, to measure the strength of the line. Since the redshift of this source is photometric, we initially left the redshift parameter in the model free to vary. From the X-ray spectral fit we obtained constraints on the redshift of $z=2.18^{+0.08}_{-0.08}$, in good agreement with the photometric estimate reported in Table \ref{tab1}. Fixing $z=2.18$ and fitting the spectrum, however, produces a very flat photon index of $\Gamma=0.70_{-0.62}^{+0.95}$, and no constraints are found on the \nh\ parameter. Fixing $\Gamma=1.8$ { yields a column density of \nh$=(4.8_{-3.0}^{+8.2})\times10^{23}$~cm$^{-2}$ with} some residuals in the hard band ($E\gtrsim2$~keV), suggesting a possible contribution from a reflection component in the spectrum. This hypothesis is supported by the high $EW$ measured for the iron line, $EW=1.2^{+1.0}_{-0.8}$~keV (rest frame), which is typical of CT AGN. { Indeed, fitting the spectrum with the Torus model results in a column density lower limit of \nh$>6\times10^{24}$~cm$^{-2}$ (see Table \ref{tab3}), i.e., in the CT regime. We tested this result also  fitting the source spectrum with a pure reflection dominated model (i.e., PEXRAV model; \citealt{magdziarz1995}) with the reflection parameter $R=-1$ fixed, plus a Gaussian line, implying column density in excess of \nh$>10^{25}$~cm$^{-2}$. This model provides a similarly good fit to the data as the previous models, yielding a spectral index of $\Gamma=2.07_{-0.43}^{+0.37}$. If XID 307N is really CT, its intrinsic luminosity would therefore be $L_{\rm 2-10~keV, intr}\approx10^{45}$~\ergs.}
\item[]\emph{XID 390N}: This $z=1.146$ quasar has a relatively strong iron K$\alpha$ line ($EW=0.38^{+0.13}_{-0.14}$~keV, rest frame, $\sigma=50$~eV fixed), which could be broad, as leaving $\sigma$ free to vary in the fit, we obtain $\sigma<556$~eV. Clear residuals suggest the presence of another emission line at $E\approx5.9$~keV (rest-frame) with $EW\approx0.28$~keV, which could be a relativistically redshifted iron line from the inner parts of the SMBH accretion disc or an outflow (e.g., \citealt{fabian2000}). Although very interesting, more investigation on the emission lines of this quasar is beyond the scopes of this paper. 

\begin{figure*}
\centerline{
\includegraphics[scale=0.83]{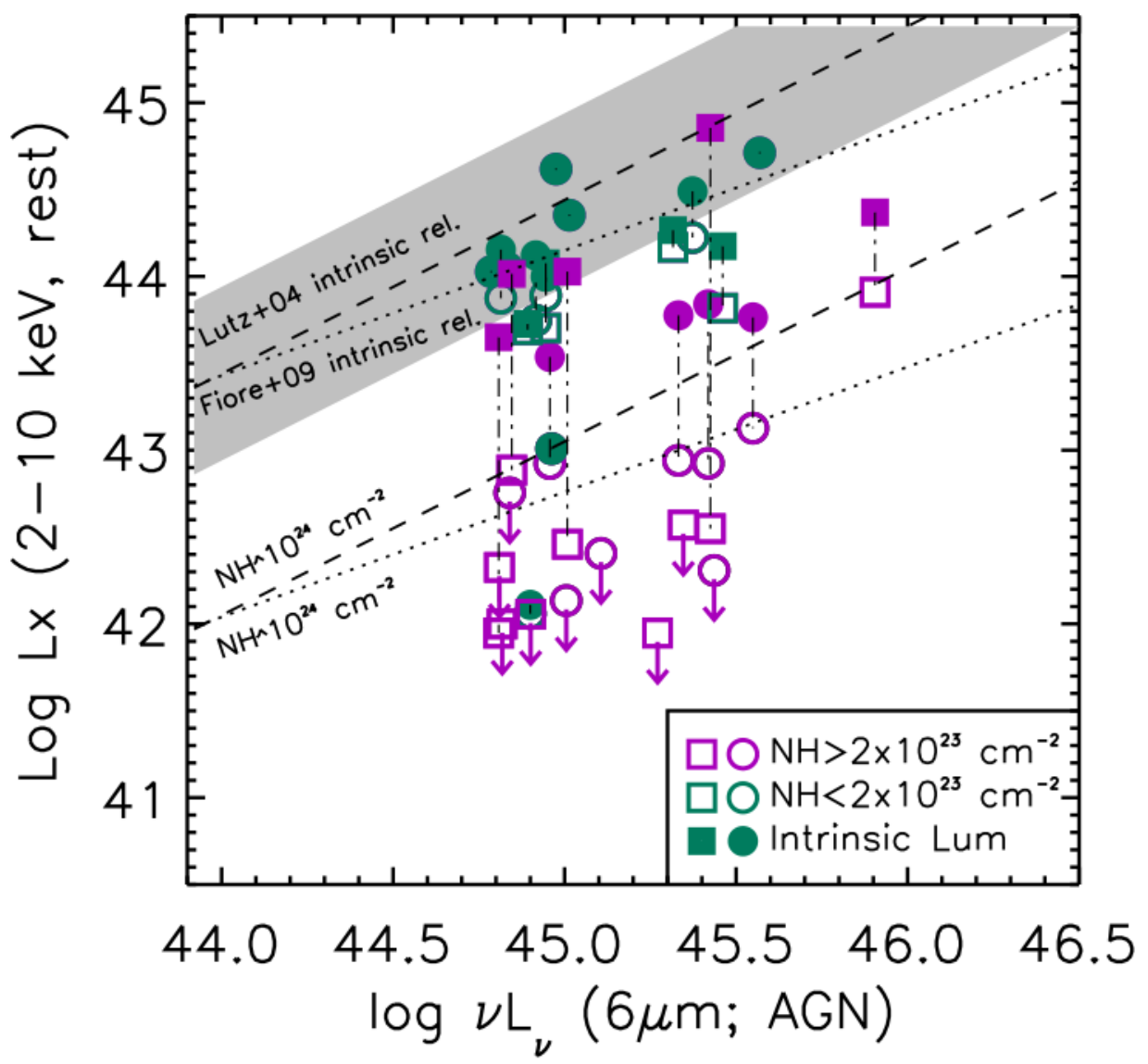}}
\caption{X-ray luminosity (2--10~keV rest-frame) vs. AGN rest-frame 6~$\mu$m luminosity (in units of \ergs) calculated from the SEDs; the observed X-ray luminosity (not corrected for absorption) is plotted with open symbols, while the intrinsic luminosity, i.e., corrected for the \nh\ measured from the X-ray spectra { (reported in Table \ref{tab2})}, is plotted with filled symbols. Unobscured and moderately obscured quasars ($N_{\rm H}<2\times10^{23}$~cm$^{-2}$) are plotted in green, while the heavily-obscured sources ($N_{\rm H}\ge2\times10^{23}$~cm$^{-2}$) are plotted in magenta. Circles and squares correspond to sources in GH-N and GH-S, respectively. The shaded region represents the scatter of the intrinsic $L_{\rm 6\mu m}-L_{\rm X}$ relation from \citet{lutz2004} (dashed line); the dotted line represents the \citet{fiore2009} relation.}
\label{fig.l6lx}
\end{figure*}

\item[]\emph{XID 149S:} This $z=2.810$ quasar has a column density \nh$\approx2\times10^{23}$~cm$^{-2}$ and a photon index $\Gamma\approx1.7$ with all the models used to fit the spectrum (i.e., model 1, PLCABS and Torus); none of the models required the addition of a soft scattered power law. Although the X-ray spectrum does not show signs of any emission line, clear residuals around $E\approx$~1.5--2~keV ($E\approx$~5.7--7.6~keV, rest frame) indicate the presence of a strong absorption feature forming a P-Cygni type profile, which indicates the presence of a high-velocity outflow (e.g., \citealt{pounds2009, tombesi2010}). Indeed, this source has been identified in the optical band as a BAL quasar (\citealt{szokoly2004}). A BAL quasar interpretation would help explaining why the intrinsic X-ray luminosity of this quasar, $L_{\rm 2-10~keV, intr}\approx2\times10^{44}$~\ergs, which is corrected for the X-ray absorption derived from the spectral fitting, is lower than that expected from the MIR luminosity $\nu L_{\rm 6 \mu m}\approx8\times10^{45}$~\ergs (see Fig. \ref{fig.l6lx}), as BAL quasars have been found to be intrinsically X-ray weak (e.g., \citealt{luo2014}). \footnote{{ The X-ray weakness of BAL quasars is typically estimated with respect to the optical band ($\alpha_{\rm OX}$; e.g., \citealt{luo2014}). However, since BAL quasars have SEDs consistent with those of typical quasars in the optical-to-IR bands (e.g., \citealt{gallagher2007}), we expect to see X-ray weakness relative to the MIR emission too.}}  
\item[]\emph{XID 435S:} Fitting this source spectrum with model 2 (see Sect. \ref{xray}) results in a column density of \nh$=1.1_{-0.7}^{+1.2}\times10^{25}$~cm$^{-2}$ (with $\Gamma=1.89^{+0.7}_{-0.6}$), which is consistent with heavily CT material. Using the PLCABS and Torus models, we obtained smaller \nh\ values of \nh$\approx (3-7)\times10^{24}$~cm$^{-2}$ ($\Gamma\approx1.9$), but still consistent with the source being CT (Table \ref{tab3}). However the spectrum does not show a clear signature of a strong Fe K$\alpha$ line. We note that, although these models are too complex for the limited counting statistics to provide tight constraints on the spectral parameters, the intrinsic $2-10$~keV X-ray luminosity corrected for the CT column densities derived from the spectral fits is in good agreement with the expectations from the MIR 6~$\mu$m AGN luminosity, according to the $L_{\rm X}-L_{\rm 6 \mu m }$ relation. XID 435S was also classified as a CT AGN also by \citet{brightman2012}.
\item[]\emph{XID 490S:} This $z=2.579$ quasar shows evidence for a strong emission line, consistent with the redshifted Fe K$\alpha$ line at $E=6.4$~keV, with an equivalent width of $EW>1$~keV (although the limited counting statistics prevent us from obtaining better constraints on the $EW$). The spectral fitting results from model 2 constrained the intrinsic column density largely in the CT regime (see Table \ref{tab1}). As in the previous case, fitting the spectrum with PLCABS or Torus models constrains the column density to lower values than that obtained from model 2 (\nh$\approx~(4-10)\times10^{24}$~cm$^{-2}$), but still consistent with heavy CT columns. Similar results were obtained by \citet{feruglio2011} and \citet{brightman2012}, who identified this source as CT.  
As for XID 435S, the intrinsic X-ray luminosity estimated by correcting for the high \nh\ obtained from the spectral fit would place this source on the $L_{\rm X}-L_{\rm 6 \mu m }$ relation (see Fig. \ref{fig.l6lx}).
\end{itemize}

\begin{table*}
\caption{Indication of CT quasar classification from various diagnostics for our CT quasars candidates.}
\begin{center}
\begin{tabular}{l c c c c c c c c}
\hline
\rule[-1.5mm]{0pt}{3ex}No & XID & Class &\nh & $EW_{\rm Fe K\alpha}$ & $L_{\rm X|6\mu m}/L_{\rm 2-10~keV}$ & $L_{\rm X|lines}$ & Notes & Refs. \\
\rule[-1.5mm]{0pt}{3ex}(1) & (2) & (3) & (4) & (5) & (6) & (7) & (8) & (9)\\
\hline
\rule[-1.5mm]{0pt}{3ex}1  & --   &          &--        &    --      & $>$522 ($>$233) & &      & 1$^{\ddagger}$  \\
\rule[-1.5mm]{0pt}{3ex}3  & 171N & CT       &  T       & \checkmark & 96 (43)         & &      & 2 \\
\rule[-1.5mm]{0pt}{3ex}6  & 243N & CT       & $\times$ & $\times$   & 270 (120)       & 45.7 & Ly$\alpha^a$ & 3\\
\rule[-1.5mm]{0pt}{3ex}9  & --   &          &  --      &   --       & $>$289 ($>$129) & &    & $^{\ddagger}$\\
\rule[-1.5mm]{0pt}{3ex}11 & 307N & { CT}& { T}  & \checkmark & 42 (19)         & &    & $^{\ddagger}$\\
\rule[-1.5mm]{0pt}{3ex}12 & --   & CT       &   --     &   --       & $>$195 ($>$87)  & 44.4 & Ly$\alpha$, C~{\sc iv}, He~{\sc ii}, C~{\sc iii}]$^b$ & 1 \\
\rule[-1.5mm]{0pt}{3ex}16 & 423N & CT       & M2, P, T &  $\times$  & 122 (54)        &    &  & 4$^{\ddagger}$ \\
\rule[-1.5mm]{0pt}{3ex}17 & --   &          &   --     &  --        & $>$47 ($>$21)   & &  &  $^{\ddagger}$\\
\rule[-1.5mm]{0pt}{3ex}23 & 320S & CT       & M2, P, T & $\times$   & 282 (126)       & &  & 5$^{\dagger}$\\ 
\rule[-1.5mm]{0pt}{3ex}24 & --   &          & --       &   --       & $>$256 ($>$114) & &  & \\
\rule[-1.5mm]{0pt}{3ex}25 & --   &          &    --    &   --       & $>$273 ($>$122) & &  & \\
\rule[-1.5mm]{0pt}{3ex}28 & --   &          &    --    &      --    & $>$820 ($>$366) & &  & $^{\ddagger}$\\
\rule[-1.5mm]{0pt}{3ex}29 & 435S & CT       & M2, P, T & $\times$   & 138 (62)        & &  He~{\sc ii}, C~{\sc iii}]$^c$ & 6 \\
\rule[-1.5mm]{0pt}{3ex}30 & 490S & CT       & M2, P, T & \checkmark & 293 (131)       & 44.5 & Ly$\alpha$, C~{\sc iv}, He~{\sc ii}, C~{\sc iii}], Ne~{\sc v}$^d$ & 6, 7$^{\ddagger}$\\
\rule[-1.5mm]{0pt}{3ex}32 & --   &          &     --   &        --  & $>$233 ($>$104) & &  & \\
\rule[-1.5mm]{0pt}{3ex}33 & --   &          &   --     &      --    & $>119$ ($>$53)  & &  & $^{\ddagger}$\\
\hline\hline
\end{tabular}
\end{center}
{NOTES: (1) Source number as in Table \ref{tab1}; (2) XID as in Table \ref{tab1}; (3) CT quasar classification: sources that we consider as secure CT quasars, based on various diagnostics explored in our analyses are marked as ``CT''; all other sources are still CT quasar candidates; (4) X-ray spectral models yielding \nh\ values consistent, within the errors, with CT absorption \nh$>1.5\times10^{24}$~cm$^{-2}$: ``M2'' refers to model 2, ``P'' to the PLCABS model and ``T'' to the Torus model (see Sects. \ref{mo1} and \ref{ct}); (5) Sources showing an iron K$\alpha$ line emission with $EW\gtrsim1$~keV; (6) Factor of suppression of the observed X-ray luminosity $L_{\rm 2-10~keV}$ compared to the intrinsic X-ray luminosity derived from the 6~$\mu$m luminosity ($L_{\rm X|6\mu m}$) using the \citet{lutz2004} relation; the values in parentheses correspond to the suppression factors estimated using the lower boundary of the intrinsic $L_{\rm X}-L_{\ 6\ \mu m}$ relation assuming a scatter of $\sim0.35$ dex; (7) Logarithm of the intrinsic X-ray luminosity derived from optical emission lines, in units of~\ergs; we adopted the average emission-line flux ratios from \citet{netzer2006} and the [O~{\sc iii}]$_{\lambda5007}$-X-ray flux ratio from \citet{mulchaey1994}; (8) List of emission lines detected in the optical spectra used to infer the intrinsic X-ray luminosity in column (7); (9) Reference of previous work where these sources are classified as CT quasars, or CT quasar candidates: $1=$ \citet{alexander2008b}, $2=$ \citet{georgantopoulos2009}, $3=$ \citet{georgantopoulos2011a}, $4=$ \citet{georgantopoulos2011b}; $5=$ \citet{brightman2014}, $6=$ \citet{brightman2012} and $7=$ \citet{feruglio2011}.
$^a$ Optical emission line information from \citet{adams2011}; $^b$  \citet{alexander2008b}; $^c$ \citet{daddi2004}; $^d$ \citet{szokoly2004} and \citet{feruglio2011}. $^{\dagger}$ This source is border-line CT in \citet{brightman2014}, but consistent with CT absorption within the errors, with \nh$=(0.5-2.9)\times10^{24}~\rm~cm^{-2}$.{ $^{\ddagger}$ These sources also have MIR-to-optical colours typical of Dust Obscured Galaxies (DOGs; see Fig. \ref{fig.f24fr} and section \ref{dogs}).}}\label{tab4}
\end{table*}%

\subsection{Obscured AGN fraction}\label{obs}

From the X-ray spectral analysis of the X-ray detected sources we found that a large fraction of the IR quasars ($\approx$67\%, 16 out of 24 sources) are absorbed in the X-ray band by column densities \nh$>10^{22}$~cm$^{-2}$. Of these sources, 9 can be classified as heavily obscured (reported in Table \ref{tab3}), with \nh$\ge2\times10^{23}$~cm$^{-2}$, of which three (XIDs 320S, 435S and 490S) have CT column densities, in excess of \nh$=1.5\times10^{24}$~cm$^{-2}$ in all the models we used (see Sect. \ref{xray}); { three} more sources (XIDs 171N{, 307N} and 423N) are consistent with being CT based on their \nh, depending on the adopted model (see Tables \ref{tab3} and \ref{tab4}), and on the detection of a strong iron K$\alpha$ emission line ($EW\gtrsim1$~keV) in the case of XID 171N { and 307N}. We therefore classify these { six} sources as CT quasars. Due to the limited numbers of counts in the spectra, however, the strong Fe K$\alpha$ line expected in CT spectra cannot be detected for all of these CT quasars.    

To assess whether our IR quasars are indeed heavily obscured, or CT, we compared the X-ray luminosity (2--10~keV, rest frame) with the AGN 6~$\mu$m luminosity measured from the best-fit SEDs (see Sect. \ref{sed}). Since the MIR band tends to be much less affected by obscuration than the X-ray band, the MIR luminosity is often used as a proxy for the intrinsic AGN luminosity (e.g., \citealt{lutz2004, alexander2008b, gandi2009}; DM13). Indeed, while unobscured sources are expected to lie on the intrinsic $L_{\rm X}-L_{\rm 6 \mu m}$ relation (e.g., \citealt{lutz2004, fiore2009, gandi2009,mateos2015}{; \citealt{stern2015}}), heavily obscured and CT sources, where the X-ray luminosity is strongly suppressed, are expected to have weak observed X-ray emission compared to the MIR emission, and therefore lie well below the $L_{\rm X}-L_{\rm 6 \mu m}$ relation. In Figure \ref{fig.l6lx} we show the rest-frame $L_{\rm 2-10~keV}$ versus $L_{\rm 6 \mu m, AGN}$ for our IR quasars; open symbols represent the observed X-ray luminosity, while the filled symbols represent the intrinsic X-ray luminosity $L_{\rm 2-10~keV, intr}$, i.e., corrected for the \nh\ derived from the X-ray spectral fitting (see Table \ref{tab2}). The unobscured and moderately obscured sources (shown in green) tend to agree with the local intrinsic $L_{\rm X}-L_{\rm 6 \mu m}$ relation, although possibly showing a flatter slope than that defined by \citet{lutz2004}, more in agreement with \citet{fiore2009}. All of the heavily-obscured IR quasars lie well below the correlation, in the region of the plot expected for CT AGN. Correcting their observed X-ray luminosity for absorption brings these sources much closer to the intrinsic $L_{\rm X}-L_{\rm 6 \mu m}$ relation, although they still lie systematically below. It is { important to note} that the \nh\ of heavily-obscured sources, obtained from spectral analyses of $E<10$~keV data alone { and with the simple parameterisation used here (i.e., model 1 and model 2)}, could be underestimated, as shown in several X-ray studies of heavily-obscured AGN performed using \ch\ or \xmmn\ data together with hard X-ray data ($E>10$~keV) from {\it NuSTAR} (e.g., \citealt{delmoro2014, gandi2014, lansbury2014}{; \citealt{lansbury2015}}). 
For the three sources securely identified as CT quasars (XIDs 320S, 435S and 490S), the intrinsic X-ray luminosity results in very good agreement with that expected from the MIR luminosity; this is also true for XID 171N { and 307N} if we correct the X-ray luminosity for the \nh\ derived from the Torus model, which gives $\log L_{\rm2-10\ keV, intr}\approx45.4$~\ergs { and $\log L_{\rm2-10\ keV, intr}\approx45.1$~\ergs, respectively}, and possibly for XID 423N considering the large scatter on the \nh\ value. This gives an extra indication that these sources are indeed CT. 

We also note that all of the X-ray undetected IR quasars (9 sources), for which we plot the X-ray luminosity upper limits, lie in the region of the $L_{\rm X}-L_{\rm 6 \mu m}$ plane expected for CT sources, and therefore can be considered as CT candidates. Two of the X-ray undetected quasars, \#1 and \#12, have been already investigated by \citet{alexander2008b} and were identified as CT on the basis of X-ray analyses and $L_{\rm X}-L_{\rm 6 \mu m}$ analyses. For \#12 the intrinsic X-ray luminosity has also been estimated from the optical emission lines (see \citealt{alexander2008b}) using the average line ratio from \citet{netzer2006} and the relation between the [O~{\sc iii}]$_{\lambda5007}$ and X-ray fluxes from \citet{mulchaey1994}, yielding $\log~L_{\rm X|lines}\approx44.4$~\ergs\ (Table \ref{tab4}), in agreement with that estimated from the MIR emission. This suggests that this source is intrinsically luminous, further confirming the CT nature of this quasar. The evidence found for \#12 also support our assumption that the X-ray undetected IR quasars are CT candidates. We therefore infer that the fraction of X-ray obscured sources (\nh$>10^{22}$~cm$^{-2}$) in our sample is $\approx$76\% (25/33 quasars, including the X-ray undetected quasars), with $\approx$55\% being heavily-obscured quasars (18/33 with \nh$>2\times10^{23}$~cm$^{-2}$) and {15} out of 33 quasars (${\approx45}$\%) are likely to be Compton thick. 
\begin{figure*}
\centerline{
\includegraphics[scale=0.6]{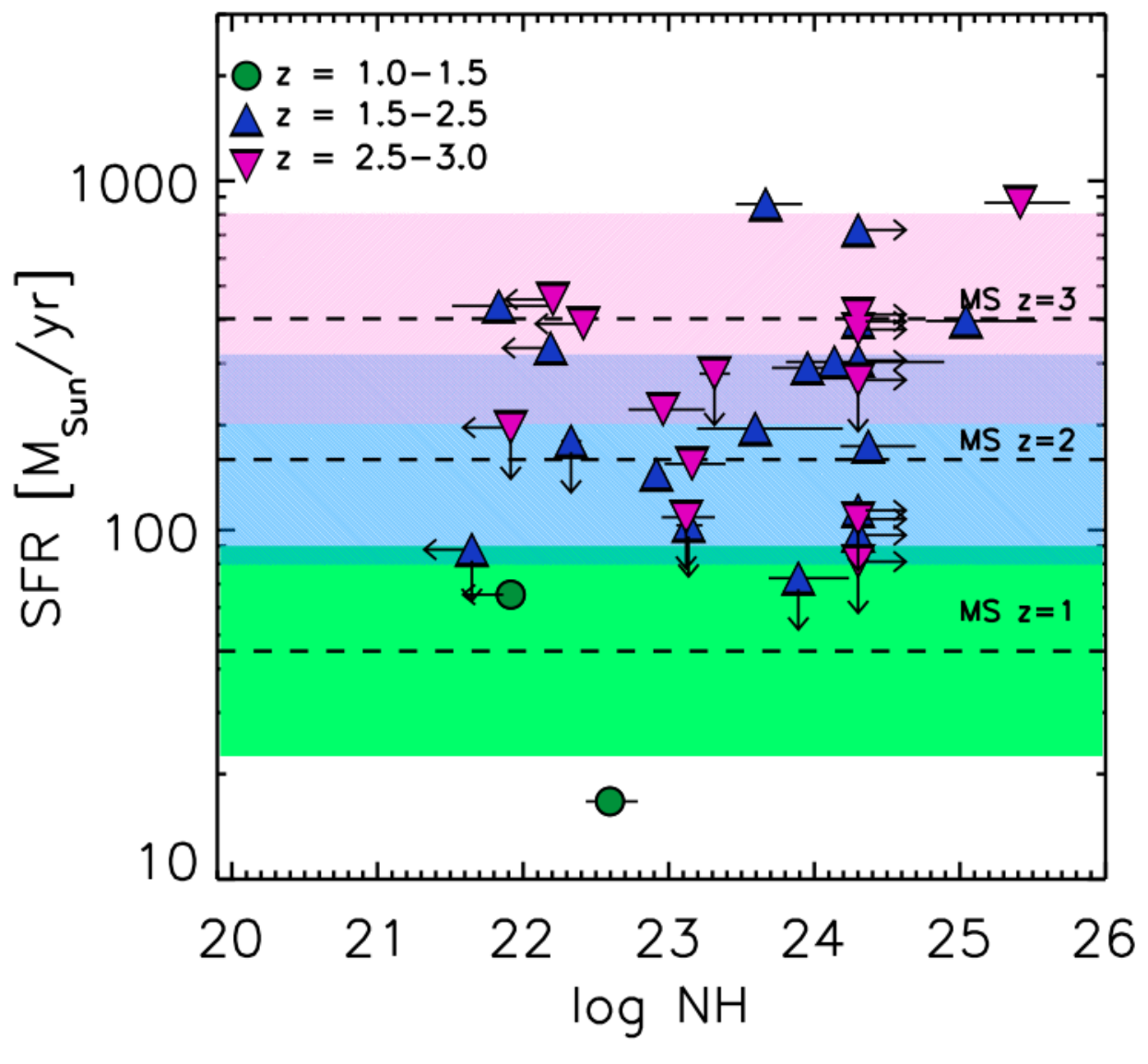}
\includegraphics[scale=0.6]{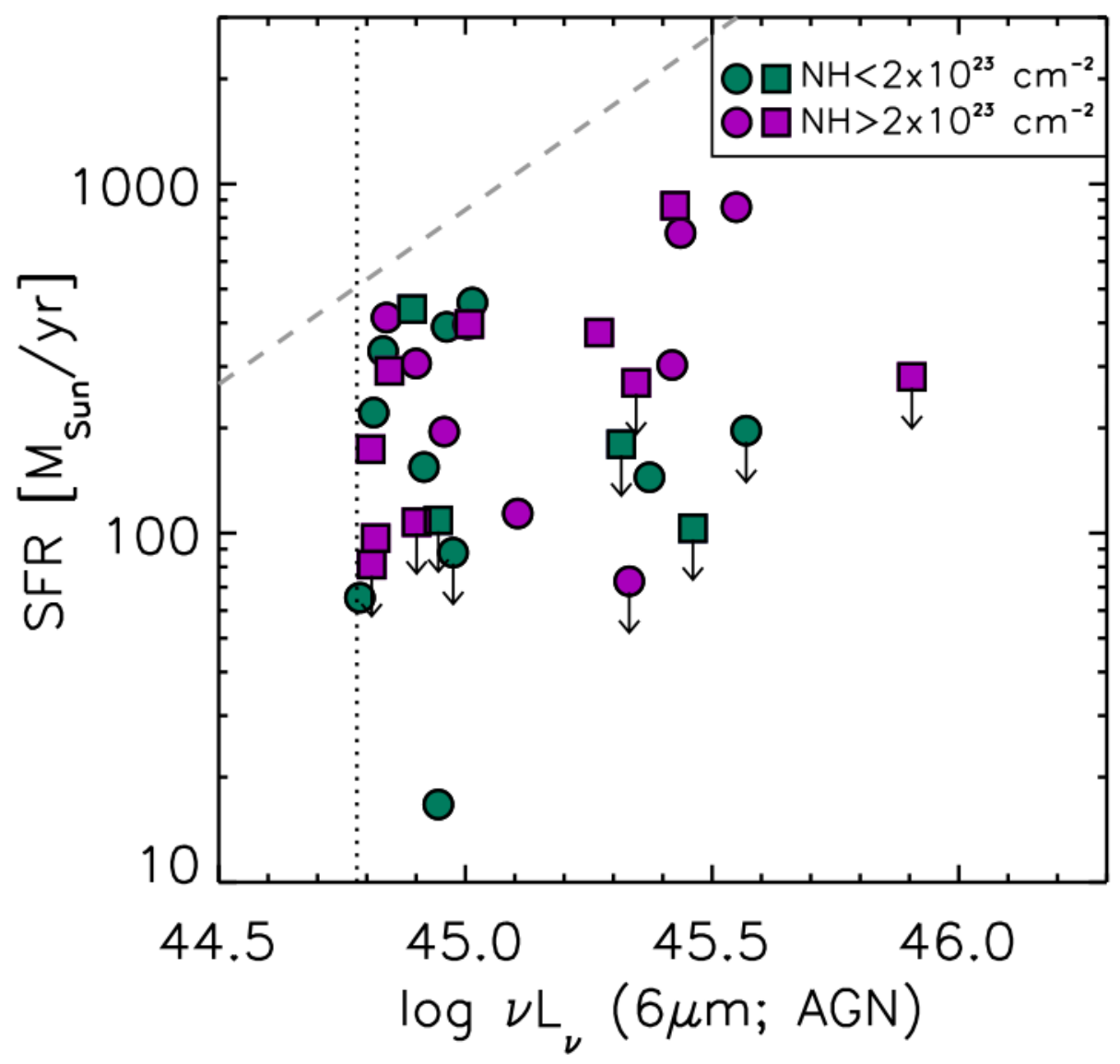}
}
\caption{Left: SFR vs X-ray column density (\nh) for the IR quasars (derived from the best-fitting of model 1 or model 2); for the X-ray undetected sources, which are CT quasar candidates, we fixed \nh\ at a lower limit of log \nh$>24.3$~cm$^{-2}$. { The sources have been divided in three redshift bins: $1.0<z<1.5$ (green circles), $1.5<z<2.5$ (blue triangles) and $2.5<z<3.0$ (magenta upside-down triangles).} The dashed lines represent the main sequence (MS) SFRs at redshift $z=1$, $z=2$ and $z=3$ (from bottom to top) for a stellar mass of $M_*=9\times10^{10}\rm\ M_{\odot}$ \citep{elbaz2011}{, which is the average stellar mass found for X-ray selected AGN at $z\approx2$ by \citet{mullaney2012}}; { the shaded areas correspond to a factor of 2 above and below the MS at each redshift ($z=1$: green, $z=2$: blue and $z=3$: pink)}. Right: SFR vs AGN 6~$\mu$m luminosity. { Symbols are the same as in Figure \ref{fig.l6lx}.} The grey dashed line indicates the limit where the AGN contributes 50\% to the total luminosity at 6~$\mu$m; our IR AGN identification approach will be incomplete for SFRs above this dashed line. The vertical dotted line at $\log\ \nu L_{\rm 6 \mu m}=44.78$~\ergs\ marks our IR quasar selection.}
\label{fig.l6sfr}
\end{figure*}

From the $L_{\rm X}-L_{\rm 6 \mu m}$ comparison it also emerged that two sources having unabsorbed X-ray spectra (XID 243N and XID 287N; see Table \ref{tab2}) have much weaker X-ray luminosity than that expected from their MIR luminosity, although for XID 287N the $L_{\rm X}/L_{\rm 6 \mu m}$ ratio is not as extreme as for XID 243N (see Fig. \ref{fig.l6lx}). XID 243N was classified as a candidate CT AGN also by \citet{georgantopoulos2011a} on the basis of its $L_{\rm X}/L_{\rm 6 \mu m}$ ratio, similarly to our analysis. The detection of a strong Ly$\alpha$ emission line in the optical spectrum, which was attributed to AGN emission rather than SF by \citet{adams2011}, allowed us to estimate the expected intrinsic X-ray luminosity (as for source \#12) $\log~L_{\rm X|lines}\approx45.7$~\ergs, calculated assuming the average AGN Ly$\alpha$ to [O~{\sc iii}]$_{\lambda5007}$ line ratio from \citet{netzer2006} and the [O~{\sc iii}]$_{\lambda5007}$ to X-ray flux ratio from \citet{mulchaey1994}. Although this luminosity is higher than that inferred from the MIR emission, possibly due to the uncertainties on the optical emission line ratios and perhaps to some contamination from SF emission to the Ly$\alpha$ flux, it supports the idea of XID 243N being intrinsically X-ray luminous; we therefore consider this source as CT (Table \ref{tab4}). In this case, if we assume that the observed $E<8$~keV spectrum is only due to scattering ($\sim$1-3\%) of the intrinsic heavily obscured emission, we estimate the intrinsic X-ray luminosity to be $\log~L_{\rm X}\approx43.6$~\ergs. This is in good agreement with the intrinsic $L_{\rm X}-L_{\rm 6 \mu m}$ relation, supporting our classification of this quasar as CT. For XID 287N no other diagnostic has been found to establish whether this source is intrinsically X-ray luminous (and therefore likely CT) or X-ray weak (e.g., \citealt{luo2014}); we thus do not include this source amongst our CT quasar candidates. 

In Table \ref{tab4} we list all the CT quasars and CT quasar candidates identified in our analyses and the various indicators that point to a CT quasar classification, i.e., primarily the high column density (\nh$>1.5\times10^{24}$~cm$^{-2}$) and/or strong iron line ($EW\gtrsim1$~keV) derived from the X-ray spectral analysis, the suppression of the observed X-ray luminosity (more than factor of $\sim30$ for CT sources) compared to the intrinsic X-ray luminosity expected from the 6~$\mu$m emission ($L_{\rm X|6\mu m}$), and other indicators found in the literature, such as the X-ray weakness compared to the intrinsic quasar luminosity derived from optical emission lines. The total fraction of CT quasars (and CT quasar candidates) in our sample is therefore $\approx${24-48}\% ({8--16} out of 33 quasars) and the total obscured quasar fraction is $\approx$79\%.  

\subsection{Star-formation rates}\label{sfr}

Using the results from our SED fitting analyses (Sect. \ref{sed}) and the X-ray spectral analyses, we explore here the connection between the quasar properties, such as their luminosity and X-ray obscuration, and the host galaxy properties, defined in terms of the SFR.
From the IR luminosity ($8-1000\ \mu$m) of the SFG component, derived from the SED fitting (see Sect. \ref{sed}), we calculated the SFR of the galaxies hosting our IR quasars. We assumed a \citet{salpeter1955} initial mass function (IMF) and used the relation from \citet{kennicutt1998}. The range of SFRs estimated for our sources is large, SFR$\approx17-866$~$\rm M_{\odot}~yr^{-1}$ (see Table \ref{tab1}). In Figure \ref{fig.l6sfr} we show the SFR versus the X-ray column density \nh\ (left) and the AGN 6~$\mu$m luminosity (right) for the IR quasars in our sample, separating the heavily obscured from the unobscured/moderately-obscured quasars (magenta and green symbols, respectively{; see Sect. \ref{obs}}). In Fig. \ref{fig.l6sfr} (left) { we divide our sample in three redshift bins: $z=1.0-1.5$, $z=1.5-2.5$ and $z=2.5-3.0$ (green circles, blue triangles and magenta upside-down triangles, respectively), and compare their SFRs with those of typical main sequence (MS) SFGs at redshift $z\approx1-3$ (e.g., \citealt{elbaz2011,mullaney2012})}. In the figure we plot the X-ray undetected sources, which are CT quasar candidates, with a column density lower limit fixed to log~\nh$>24.3$~cm$^{-2}$.   
{ On average there is good agreements with the SFRs of our IR quasars and the MS in each redshift bin, however, at $z=1.5-2.5$ there is a significant fraction of sources with SFRs more that a factor of 2 above the MS (6/18 sources in this redshift bin, i.e., $\approx33$\%), which could be classified as starburst galaxies; these sources also tend have higher X-ray absorption (\nh$>2\times10^{23}$~cm$^{-2}$). On the other hand at $z=2.5-3.0$ only one (heavily-obscured) quasar out of 13 has SFR typical of starburst galaxies at similar redshift, while a significant fraction of our IR quasars (5/13 sources, $\approx$38\%) have SFRs below the MS. This possibly suggests a bias in our sample, as we are likely to miss quasars with very high SFRs (see also Fig. \ref{fig.l6sfr}, right, and Sect. \ref{irquasar}). We caution, however, that given the small number of sources in each redshift bin, these results are only tentative.}

Overall, we do not find any significant correlation with the intrinsic AGN luminosity, $L_{\rm 6 \mu m, AGN}$, although the range in luminosity for our IR quasars is { relatively narrow}; this is in agreement with other studies by \citet{harrison2012b} and \citet{stanley2015}. { Interestingly, on average we find that} the most obscured quasars have higher SFRs than the unobscured quasars (see also \citealt{chen2015}). However, as the number of sources in our sample is small, this difference is mainly driven by three heavily-obscured sources with very high SFRs { ($\approx$700-900~$\rm M_{\odot}~yr^{-1}$, $\sim$16\% of the heavily-obscured quasars) that are not representative of the average IR quasar host population,} for which there seems to be no significant relation between SFR and X-ray obscuration (e.g. \citealt{lutz2010,rosario2012,rovilos2012}).

{ From the measured SFRs we calculate the expected X-ray emission at 2-10~keV (rest frame; $L_{\rm X,SF}$) produced by star formation using the \citet{lehmer2010} relation between SFR and rest-frame 2--10~keV luminosity (equation 4 in their Table 4). The values are reported in Table \ref{tab1}. We also estimate what fraction of the observed X-ray luminosity (at 2-10~keV, rest frame) could therefore be due to star formation (see Table \ref{tab1}). For the majority of the sources ($\approx$82\%) the contribution from star formation is negligible, of the order of $<$10\%. However, for $\approx$18\% of the sources more that 10\% of the observed X-ray luminosity could be due to star formation and for some of them (\#1, 6, 9 and 28) this contribution can reach up to $\approx$20-30\%. All of these quasars are candidates CT according to our analyses, of which 3 out of 4 are X-ray undetected. The relatively high contribution from star formation to the 2-10~keV luminosity (or upper limit) supports the idea that in these quasars we do not detect the intrinsic AGN emission at X-ray energies, which is likely to be very heavily absorbed, as discussed in section \ref{xsrc} for XID 243N (\#6). }  

\subsection{Host galaxy interactions}\label{morph}

In order to test the major merger evolutionary scenario for quasars, which predicts that most of the accretion onto the SMBH happens in heavily obscured environments, due to large amounts of gas and dust driven to the centre of galaxies during the merger (e.g., \citealt{sanders1988, dimatteo2005, hopkins2006a}; see also \citealt{alexander2012}, for a review), we investigate here the connection between the X-ray obscuration and the host galaxy morphology of our IR quasars. In particular, we aim to explore whether the most obscured IR quasars are typically found in interacting/merging systems.  

We used the high resolution HST images to classify the galaxies in terms of disturbance or distortion of their morphology. The sources were visually classified mainly using the $H$-band images, however the supplementary information from the $V$, $z$, and $J$ bands were also used (see Sect. \ref{odata}). Following the visual classification method described in Section 4 of \citet{kocevski2012}, the galaxies were initially classified as: 1) ``mergers'', where strong signatures of distortions and/or multiple nuclei in a single coalescing system were visible; 2) ``interactions'', i.e., systems with two distinct galaxies clearly showing interaction features, like tidal streams; 3) ``distorted'', individual galaxies with no visible companion, but showing asymmetric morphology or distortions; 4) ``double nuclei'', i.e., galaxies with multiple nuclei in a single coalesced system; 5) ``close pairs'', where two (or more) galaxies lie within $\sim$1.5$''$--2.0$''$ (corresponding to $\sim$13--17~kpc at $z\approx2$), although no signs of interactions are visible (which could mean the pairs are just line-of-sight alignments); 6) ``not-disturbed'', single galaxies with no distortion/interaction signatures.
We adopted the same classification method as in \citet{kocevski2012} in order to allow direct comparison between our results and those presented for their galaxy and AGN samples, which we consider here as control samples for our IR quasars (as they span similar $z$ ranges). 
\begin{figure}
\centerline{
\includegraphics[scale=0.24]{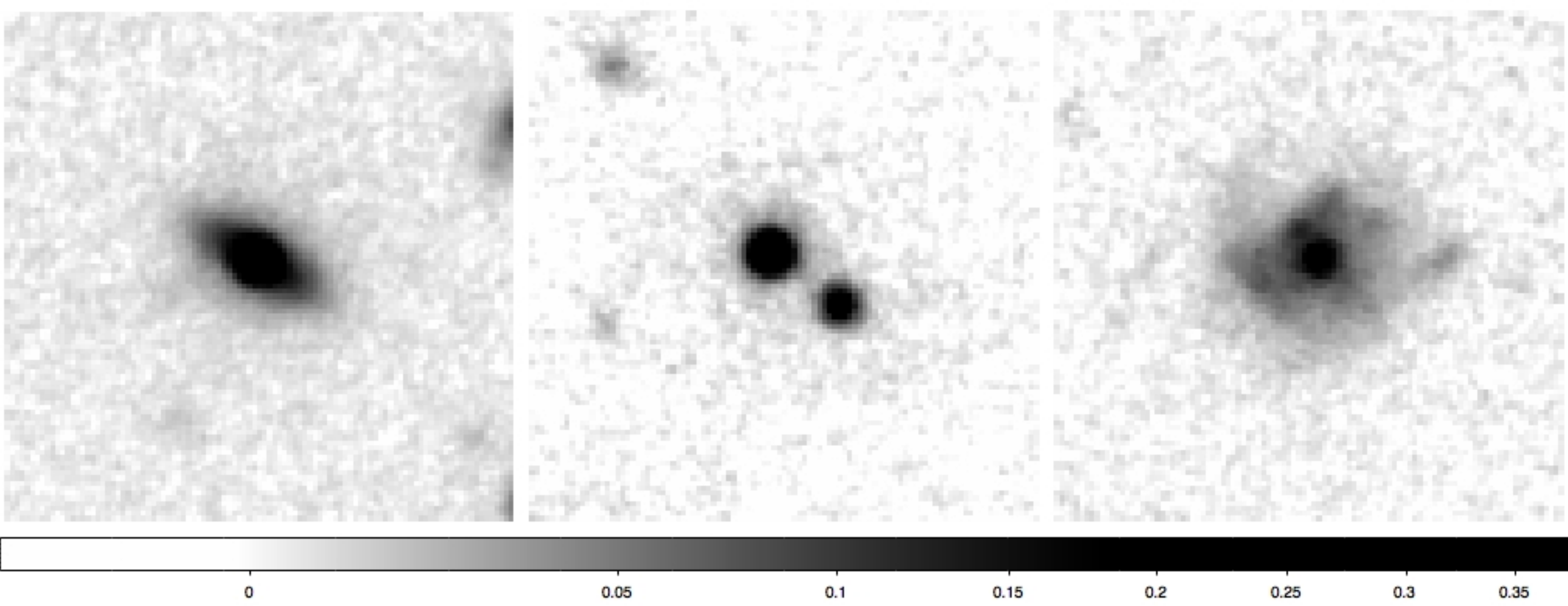}}
\caption{HST F160W ($H$ band) image cutouts ($6''\times6''$) for three of our IR quasars showing examples of an ``undisturbed'' galaxy (\#14, left), a ``companion'' class (\#5, centre) and a ``disturbed'' galaxy (\#24, right). The source numbers are those reported in column (1) in Table \ref{tab1}.}
\label{fig.hima}
\end{figure}

Since the number of sources in our sample, and therefore in each class of objects, is small, we grouped the disturbance classes listed above in two main groups, following \citet{kocevski2012}: a) \emph{``Disturbed''}, which include all galaxies showing any indication of distortion or interaction, i.e., classes 1), 2) 3) and 4); this group corresponds to the \emph{Disturbed II} class in \citet{kocevski2012}; b) \emph{``Undisturbed''}, i.e., all sources in classes 5) and 6). Although class 5), hereafter defined as \emph{``Companion''} group, is a sub-class of the undisturbed group, we also consider it separately, as despite the fact that these close pairs do not show visible signs of distortions/interactions they might still be associated. Examples of these three groups are shown in Figure \ref{fig.hima}. To test the robustness of our classification method, we compare our disturbance classification for the IR quasars in the GH-S field with the visual classification performed by the CANDELS team (e.g., \citealt{kartaltepe2012}), described in Section 2.1.2 of \citet{rosario2015}.\footnote{The CANDELS visual classification is an ongoing program currently including morphological and disturbance classifications for all the galaxies with $H<24.5$ mag in the GOODS-S and UDS fields, and will soon provide classifications in all the remaining CANDELS fields. In their catalogue every source is inspected by at least 5 classifiers who assigned a number to each source as a measure of their disturbance and interaction/merger level (or ``interaction metric'', IM).} For details on the classification method and the reliability of the classification and classifiers we refer to \citet{kartaltepe2012}. According to the CANDELS visual classification, sources are defined as ``disturbed'' if at least 2/3 of the classifiers identified them as such, ``interacting'' if the interaction metric (IM) is 0.2$<$IM$<$0.5 and ``mergers'' if IM$\ge$0.5 (see \citealt{rosario2015}). We find an agreement of 86\% between our classification and the CANDELS visual classification of the same sources in GH-S; we therefore consider our disturbance assessment of the full IR quasar sample as reliable.
\begin{figure}
\centerline{
\includegraphics[scale=0.62]{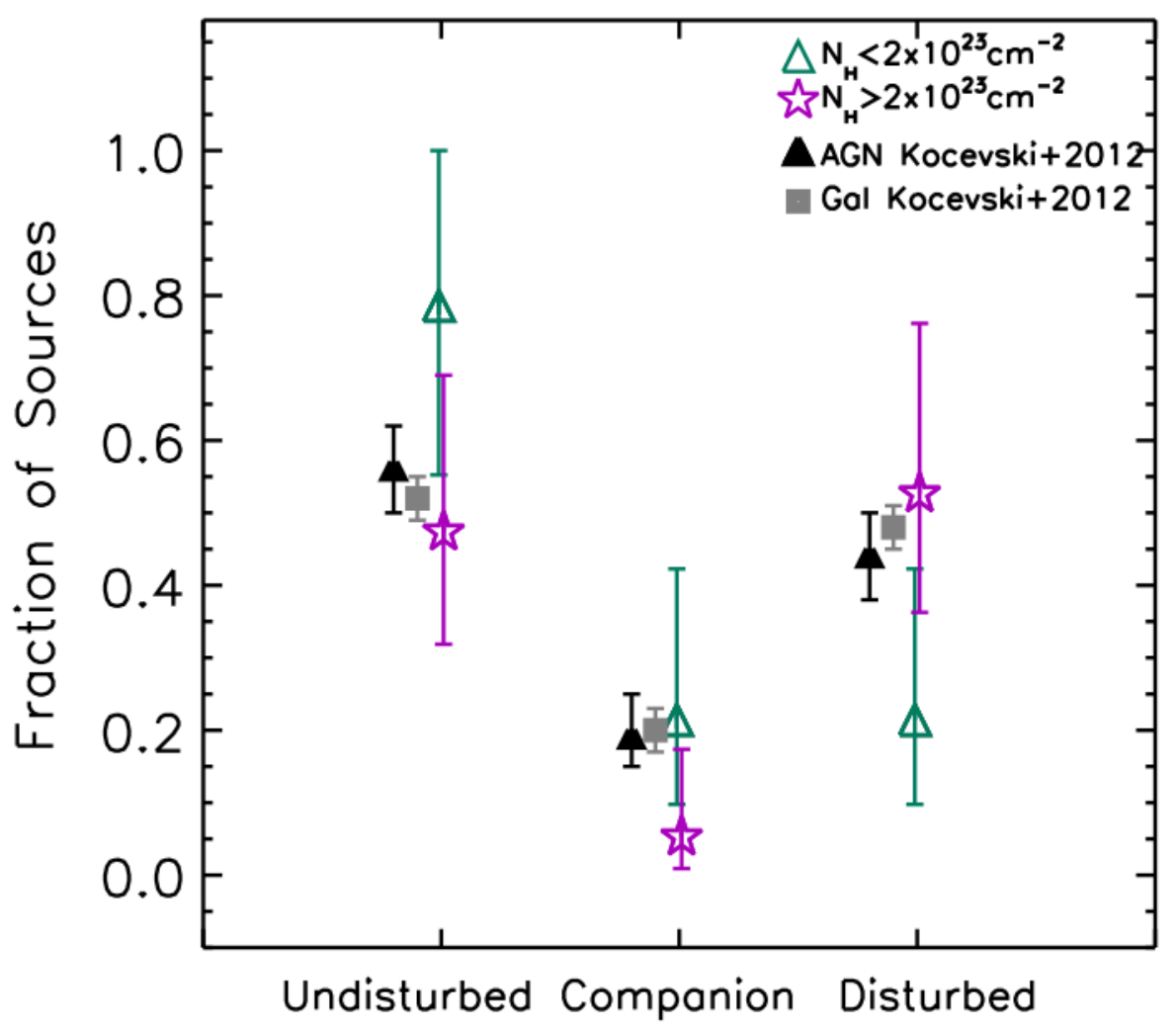}}
\caption{Fraction of IR quasar hosts with various disturbance classes, classified using HST images as described in Sect. \ref{morph}. The fractions and 1$\sigma$ uncertainties for the unobscured/moderately-obscured quasar hosts (\nh$<2\times10^{23}$~cm$^{-2}$) are plotted as green triangles and for the heavily-obscured quasars (\nh$\ge2\times10^{23}$~cm$^{-2}$) as magenta stars {(see Sect. \ref{obs})}. The classes are divided into: ``undisturbed'', ``companion'' (which are a sub-class of the undisturbed galaxies) and ``disturbed''. For comparison the fractions for $z\approx2$ AGN and non-AGN samples from \citet{kocevski2012} are also shown (filled black triangles and grey squares, respectively). The disturbance fractions for our IR quasars are in good agreement with \citet{kocevski2012}. The unobscured/moderately-obscured quasars reside preferentially in undisturbed systems, while the heavily-obscured quasars are equally found in disturbed and undisturbed systems.}
\label{fig.morph}
\end{figure}

In Figure \ref{fig.morph} we plot the fraction of IR quasar host galaxies belonging to each of the three disturbance groups described above (disturbed, undisturbed and companion). We separated the unobscured/moderately-obscured sources (green) and the heavily obscured sources (magenta{; see Sect. \ref{obs}}). The 1$\sigma$ errors in the plot are calculated following \citet{gehrels1986}. The total fraction of IR quasars residing in undisturbed galaxies is $\approx60$\%, while that of the sources in disturbed/interacting systems is $\approx40$\%. These fractions are in good agreement with those found by \citet{kocevski2012} for samples of AGN ($\approx$56\% undisturbed and $\approx$44\% disturbed) and non-AGN hosting galaxies ($\approx$52\% undisturbed and $\approx$48\% disturbed) at similar redshifts to our sample. This is true also for the fraction of sources showing a close companion ($\approx$12\%). Separating the sources between heavily obscured and unobscured/moderately-obscured quasars, we find suggestive evidence that the large majority of the unobscured quasars reside in undisturbed galaxies ($\approx$80\%, as opposed to $\approx$47\% for the heavily-obscured quasars), with only $\approx$20\% having visible signs of interactions; this fraction seems to be higher ($\approx$53\%) amongst the heavily-obscured quasars. { Consistent results were found for a sample of X-ray selected heavily-obscured AGN in the COSMOS field by \citet{lanzuisi2015}.} The drop in the number of disturbed galaxies amongst the unobscured/moderately-obscured quasars might partly be due to the fact that the bright unobscured nucleus might impair the detection of faint signs of interactions, such as tidal features; we note that particular care has been applied to the classification of these sources to minimise this possible bias. Although the errors on the fractions of disturbed and undisturbed galaxies are relatively large, we find that the difference between unobscured/moderately-obscured quasars and heavily-obscured quasars is significant at 90\% confidence level (Fisher exact probability test, $P=0.087$). However, larger samples are necessary to reliably assess these differences at higher confidence levels. 

\section{Discussion}\label{discus}

We have investigated the X-ray properties of a sample of 33 MIR selected quasars with intrinsic AGN luminosity $\nu L_{\rm 6 \mu m}>6\times10^{44}$~\ergs\ at redshift $z\approx1-3$. While at MIR wavelengths these quasars are very luminous, the majority have low X-ray luminosities, and a significant fraction of them ($\sim$30\%) are undetected in the deep \ch\ data available in the GH fields, suggesting they are heavily obscured in the X-ray band. Indeed, the X-ray spectral analysis and the comparison between the X-ray and MIR luminosities indicate that the vast majority of these IR quasars are obscured (up to $\sim79$\%, 26 out of 33 sources), of which a large fraction are candidates to be CT (up to ${ \sim48}$\%, i.e., { 16} out of 33 quasars). The fractions of obscured and potentially CT quasars are much higher than those typically found for optically and X-ray selected quasar samples (see Sect. \ref{frac}). We have reliably identified { 8} of these sources as CT from the X-ray spectral analyses (\nh\ and strong Fe K$\alpha$ line), although some ambiguity still remains for { three} of them (XIDs 171N{, 307N} and 423N) due to the large uncertainties on the estimated \nh, from the suppression of the X-ray luminosity (a factor of $\approx30-1000$) compared to the 6~$\mu$m luminosity derived from the SED analyses and from the optical emission line luminosity. Further diagnostics are needed for the remaining 8 CT quasar candidates to confirm their nature. 
Eight of our CT quasar candidates have also been classified as CT (or CT candidates) in previous studies (see Table \ref{tab4}, for reference), however, the remainders have never been identified to date. Our results therefore indicate that there is a large population of obscured and heavily-obscured quasars at $z\approx2$ and confirm that the selection of sources in the MIR band allows us to find even the most obscured, CT quasars, that are missing in the optical and X-ray bands.

\subsection{Obscured quasar fraction: comparison with previous studies}\label{frac}

Several studies conducted in the optical and X-ray bands have shown that the fraction of obscured AGN decreases with luminosity (e.g., \citealt{lawrence1982, ueda2003, steffen2003, hasinger2004, simpson2005,lafranca2005, treister2005, hasinger2008, ueda2014}). For instance, \citet{hasinger2008} found that the obscured fraction ($f_{\rm Cthin}$), defined as the number of obscured, Compton-thin AGN (Cthin; \nh$=10^{22}-10^{24}$~cm$^{-2}$) over the total with \nh$<10^{24}$~cm$^{-2}$, amongst quasars is more than a factor of 2 lower ($f_{\rm Cthin}<$25\% at $L_{\rm X}>10^{44}$~\ergs) than that for the less-powerful AGN ($f_{\rm Cthin}\approx$50\% at $L_{\rm X}=10^{43}$~\ergs). This trend is also seen when selection biases are accounted for (due to the fact that unobscured bright quasars are easier to detect out to larger comoving volumes, e.g., \citealt{dellaceca2008}). The obscured AGN fraction seems also to evolve with redshift, becoming larger at high redshift (e.g., \citealt{lafranca2005,ballantyne2006, treister2006,hasinger2008,ueda2014}). The trend of decreasing $f_{\rm Cthin}$ at high luminosity, however, is found at all redshifts (e.g., \citealt{hasinger2008,ueda2014}).

To compare the obscured fraction in our sample with those found in previous studies we consider here the Compton-thin quasars only, i.e., excluding sources with \nh$>10^{24}$~cm$^{-2}$. Amongst our IR quasars $f_{\rm Cthin}={59\%-72}$\%, where the range is determined by the lower and upper limit on the number of candidate CT quasars in our sample. These fractions are higher than those found by \citet{hasinger2008} for X-ray and optical selected $L_{\rm X}\approx10^{44}$~\ergs\ quasars at $z=1-3$, $f_{\rm Cthin}\approx40\%-55$\%. However, we find good agreement with the \citet{ueda2014} estimates in the same redshift range, $f_{\rm Cthin}\approx60\%-65$\% at $L_{\rm X}\approx10^{44}$~\ergs, where they apply a correction for detection biases to the observed X-ray absorbed AGN fraction. 

Taking into account the CT sources in our sample the obscured quasar fraction reaches $\approx79$\% (see Sect. \ref{obs}), with CT quasars being up to $\sim${62}\% of the total obscured quasars. This fraction is consistent with those found for Seyfert galaxies ($L_{\rm X}<10^{44}$~\ergs) in the local universe, where $\sim$50\% of the AGN population shows CT absorption (e.g., \citealt{risaliti1999, guainazzi2005, tozzi2006}), possibly suggesting that the relative ratio of Cthin and CT sources remains constant across all redshifts and luminosities. 
Our results also support the assumptions of the CXB synthesis models by \citet{gilli2007} and \citet{ueda2014} of an equal number of Cthin and CT AGN across all luminosities and redshifts. However, we note that different population synthesis models (e.g., \citealt{ballantyne2006,treister2009}) use different assumptions on the relative numbers of Cthin and CT AGN (see Table 6 of \citealt{ueda2014}). \citet{buchner2015} found that obscured AGN (\nh$>10^{22}$~cm$^{-2}$) at $\log L_{\rm X}>43.2$~\ergs\ contribute $\approx75$\% of the luminosity density at $z=1-3$, with CT AGN and obscured Cthin AGN contributing in equal fractions, while unobscured AGN only making up $\approx25$\% of the luminosity density. We found consistent results for quasars ($L_{\rm X}>10^{44}$~\ergs).
 
In the context of galaxy evolution, where the most massive galaxies, hosts of the most luminous quasars, form earlier and evolve more rapidly than the less luminous systems (cosmic ``downsizing'', e.g., \citealt{cowie1996, franceschini1999}), a decreasing fraction of obscured sources with luminosity can be explained with the most luminous AGN being more efficient than their less luminous analogues in consuming their surrounding gas and blowing off their dusty cocoon (through AGN feedback; e.g., \citealt{hopkins2006a,menci2008}) becoming unobscured, bright quasars. This means that the probability of seeing obscured high-luminosity quasars is smaller than for lower luminosity AGN, as the fast accreting SMBHs spend a shorter time in this ``obscured phase''. However, at high redshifts we might expect to see more of these bright obscured quasars as they have not yet expelled or consumed all their surrounding material. { Indeed, \citet{assef2015} found that the space density of very luminous obscured quasars at high redshift ($2<z<4$) is comparable to that of unobscured quasars of similar luminosities.} In this scenario it is perhaps not surprising to find such a high fraction of obscured quasars ($\sim80$\%, including CT sources) in our sample at $z\approx2$. In fact, there is indication that CT AGN have a similar evolution with redshift to Cthin AGN (e.g., \citealt{brightman2012}; but see also \citealt{buchner2015}). The selection in the IR band tends to find the most obscured, CT AGN more efficiently than in the optical and X-ray bands, and previous studies of MIR selected AGN have indeed found high fractions of obscured and CT sources, consistent with our results (e.g., \citealt{maiolino2007, treister2008, fiore2008, georgantopoulos2011b,stern2014}). 
\begin{figure}
\centerline{
\includegraphics[scale=0.6]{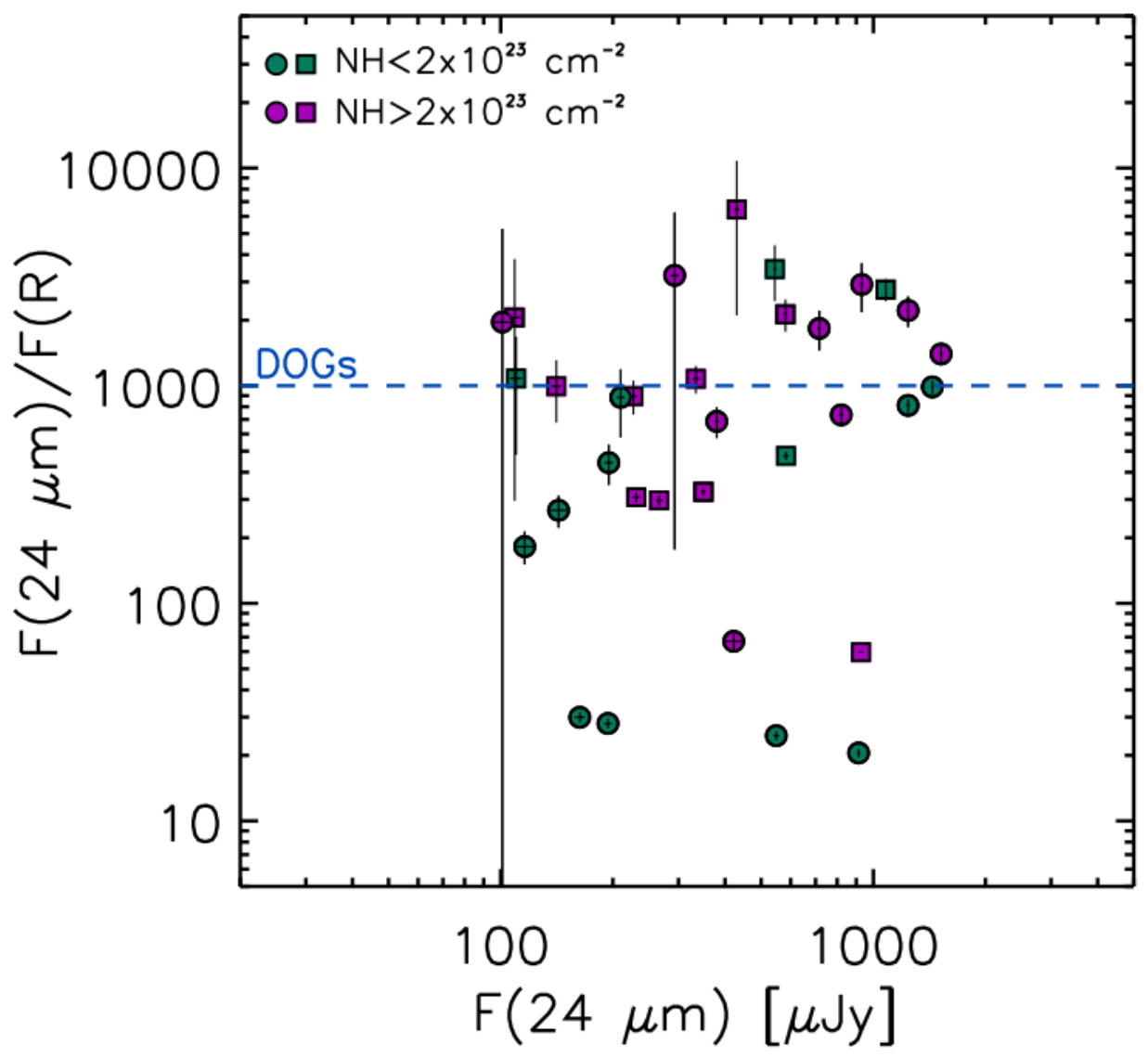}}
\caption{{ Mid-infrared to optical flux ratio, $\rm F(24\ \mu m)/F(R)$, vs the \spz\ 24~$\mu$m flux density in units of $\mu$Jy for our IR quasar sample. The sources in GH-N are shown as circles and the sources in GH-S as squares; heavily-obscured quasars are represented in magenta, while the moderately-obscured/unobscured quasars are in green (as in Fig. \ref{fig.l6lx}). The dashed line marks the flux ratio defining Dust Obscured galaxies (DOGs, $\rm F(24\ \mu m)/F(R)>1000$; e.g., \citealt{fiore2008}). The fraction of IR quasars in our sample that can be classified as DOGs is $39_{-22}^{+24}$\%.}}
\label{fig.f24fr}
\end{figure}

\subsection{Comparison with the Dust Obscured Galaxy population}\label{dogs}

X-ray obscured AGN have often been found to show extreme MIR-to-optical colours ($\rm F(24\ \mu m)/F(R)$; where $\rm F(24\ \mu m)$ is the flux density at 24~$\mu$m (observed frame) and $\rm F(R)$ is the flux density in the $R$ band), due to strong reddening typically affecting the optical bands (e.g., \citealt{fiore2009,georgantopoulos2009}). \citet{fiore2008} identified a MIR-to-optical colour selection, $\rm F(24\ \mu m)/F(R)>1000$, as an efficient way to select heavily obscured, CT AGN, finding a large fraction ($\approx$80\%) of such sources in their sample. Objects with these extreme MIR-to-optical colours have often been called Dust Obscured Galaxies (DOGs; e.g., \citealt{dey2008,pope2008a}).

To investigate how our IR quasar selection, and our obscured quasar fraction compare to the DOG population, we plot the flux density ratio $\rm F(24\ \mu m)/F(R)$ versus the flux density at 24~$\mu$m for all our IR quasars (Figure \ref{fig.f24fr}). The $R$ fluxes for the GH-N sources were derived from the Subaru Suprime-Cam $R$-band magnitudes \citep{capak2004}, while for the GH-S sources they are extrapolated from the HST-ACS $v$ and $i$-band magnitudes (\citealt{giavalisco2004}).  
We find that a relatively small fraction of $39_{-22}^{+24}$\% (the errors on the fraction are estimated from the errors in the $\rm F(24\ \mu m)/F(R)$ ratio and Poisson uncertainty) of our IR quasars show the extreme colours typical of DOGs. The fraction of heavily obscured quasars amongst them (10/13 sources; $\approx77$\%) is in good agreement with that found by \citet{fiore2008}. { Of these sources, 8 are amongst our CT quasar candidates (see Table \ref{tab4}).} However, almost half of the quasars that we identified as heavily obscured in our sample would be missed by the DOG colour selection (9/19; $\approx47$\%){, including $\approx$50\% of our CT quasar candidates (8/16 sources).\footnote{ We note that the optical and near-IR photometry of the source \#6 (XID 243N) is contaminated by a nearby low-$z$ galaxy (see also note in Table \ref{tab1}), and therefore its $\rm F(24\ \mu m)/F(R)$ ratio, which results to be well below the DOG selection threshold, might not be reliable.}} This indicates that { the DOG colour selection is highly incomplete in identifying CT quasars and}, even though some of the most heavily obscured quasars are likely to have extreme $\rm F(24\ \mu m)/F(R)$ colours, this is not true for all of them. Indeed, as discussed in section \ref{irquasar}, the colour selections can be strongly affected by contamination from the host galaxy emission and by the k-correction, as the source redshift is not accounted for in these methods, and therefore cannot provide a clean and complete selection of sources. For our sources we do not find any significant relation between the $\rm F(24\ \mu m)/F(R)$ ratio and the redshift or the SFRs; further investigation on which of these effects have the biggest impact on the DOG colour selection, however, is beyond the scopes of this paper. 
 
\subsection{Space density of CT quasars}\label{space}

Considering the full IR quasar sample, we calculate the sky density of our sources ($\nu L_{\rm 6 \mu m}>6\times10^{44}$~\ergs at $z=1-3$) within the GH fields (North and South; $\sim260$ arcmin$^2$ in total) to be $\approx$460 deg$^{-2}$. 
CT quasars constitute up to ${\approx48}$\% of our sources, i.e., up to ${\approx62}\%$ of the obscured quasars in our sample (26 objects with \nh$>10^{22}$~cm$^{-2}$). 
Calculating the space density of CT quasars (and CT candidates) in our sample ($\sim${8-16} sources) in the comoving volume between $z=1-3$ we find $\Phi={(6.7\pm2.2)}\times10^{-6}$~Mpc$^{-3}$, where the errors are estimated from the lower and upper limits on the number of CT quasars in our sample. 

\begin{figure}
\centerline{
\includegraphics[scale=0.65]{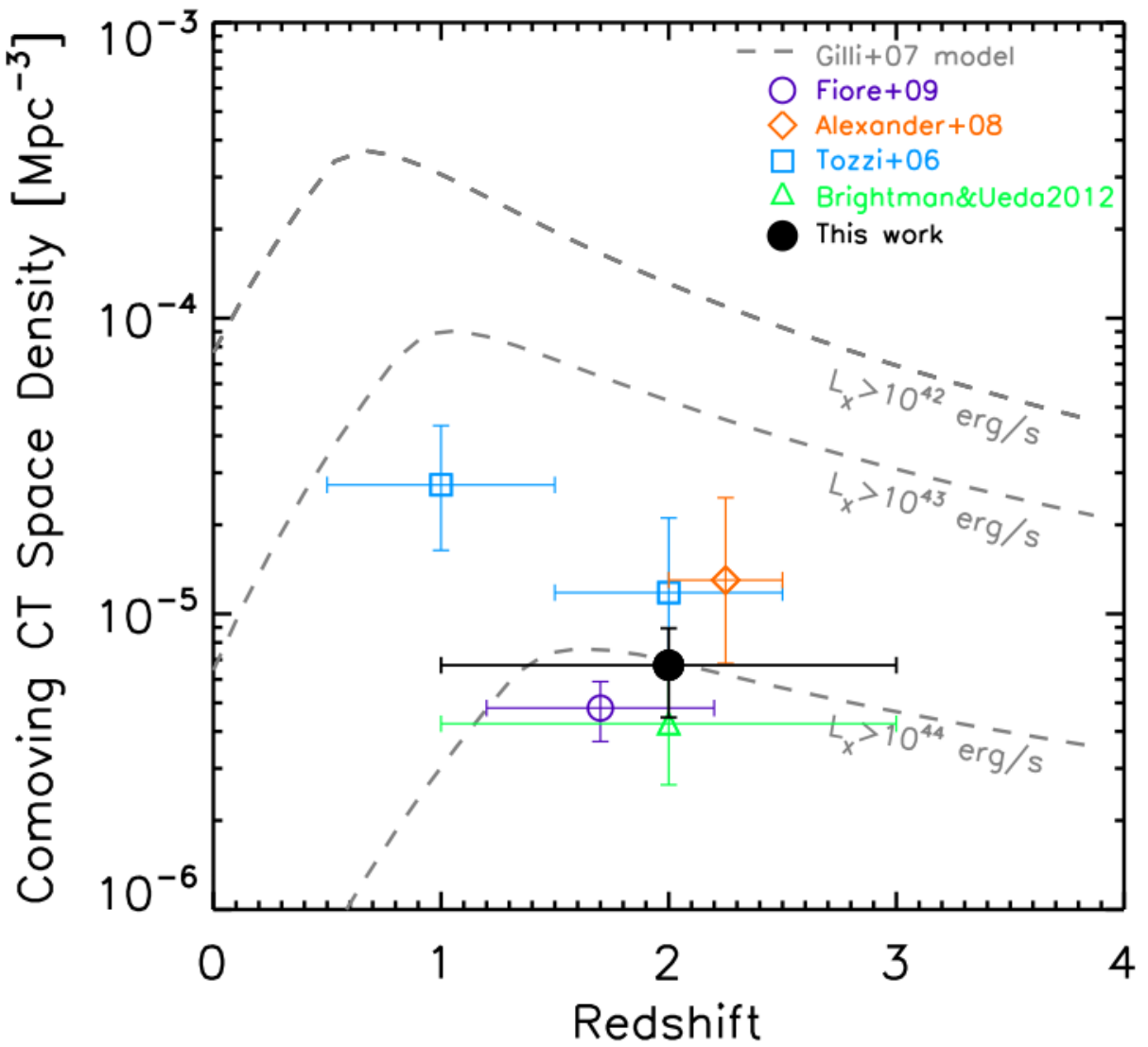}}
\caption{Space density of CT quasars at redshift $z=1-3$ estimated from our IR quasar sample within the GH fields (black circle). The grey dashed lines are the space density predictions from the \citet{gilli2007} model for AGN with $L_{\rm X}>10^{44}$~\ergs, $L_{\rm X}>10^{43}$~\ergs\ and $L_{\rm X}>10^{42}$~\ergs\ (from bottom to top). Results from various previous studies are also shown: the space density of quasars with $L_{\rm X}>10^{44}$~\ergs\ at $z\approx2$ from \citet{brightman2012} (green triangle), \citet{fiore2009} (purple circle), \citet{alexander2008b} (orange diamond)  and \citet{tozzi2006} (blue squares). We also show the space density of AGN with $L_{\rm X}>10^{43}$~\ergs\ at $z\approx1$ from \citet{tozzi2006}.}
\label{fig.ctdens}
\end{figure}

In Figure \ref{fig.ctdens} we plot the comoving CT quasar space density estimated in this work and compare it with those found in previous studies. 
Our results are consistent with the space density of $L_{\rm X}=10^{44}-10^{45}$~\ergs\ (intrinsic) CT quasars at $z\approx1.2-2.2$ found by \citet{fiore2009}, who based their estimates on IR selected sources in the COSMOS field. \citet{alexander2008b} calculated the volume density of CT quasars at $z\approx2-2.5$ in the CDF-N based on 4 sources identified in the X-ray and IR bands; their estimate is somewhat higher than ours, but it is in broad agreement given their large uncertainties. Our estimate is in good agreement with the CT quasars space density predicted by the \citet{gilli2007} model for intrinsic $L_{\rm X}>10^{44}$~\ergs\ at $z\approx2$ and also with the values found by \citet{tozzi2006} and \citet{brightman2012}, from X-ray spectral analyses in the CDF-S (Fig. \ref{fig.ctdens}). 

For a more direct comparison with the X-ray studies of \citet{tozzi2006} and \citet{brightman2012}, we limit our sample here to the CT quasars (and candidates) identified in the GH-S field (8 sources, three of which are detected in the X-rays). The space density estimated by \citet{tozzi2006} for CT AGN at $z\approx2$, is based upon 4 CT AGN identified from X-ray spectral analysis in the central 95~arcmin$^{2}$ in CDF-S (1 Ms). \citet{brightman2012} identified $\sim10$ CT AGN in the CDF-S ($\approx460$~arcmin$^{2}$) at $z\approx1-3$ and intrinsic $L_{\rm X}>10^{44}$~\ergs, yielding a similar volume density to \citet{tozzi2006} and to ours. However, it is important to note that there is no overlap between the CT AGN identified by \citet{tozzi2006} and those identified in this work through spectral analysis (i.e., the three X-ray detected sources in GH-S). Two of our three X-ray detected CT quasars were undetected in the 1 Ms \ch\ observations analysed in \citet{tozzi2006}, while the third source was mis-classified as a low-luminosity, unobscured AGN. On the other hand, all the CT AGN identified by \citet{tozzi2006} have lower MIR AGN luminosity than our IR quasars (although still consistent with the scatter of the $L_{\rm X}-L_{\rm 6 \mu m}$ relation), and therefore are not selected in our sample, as we are only probing the most luminous IR quasars. Two out of the three X-ray detected CT quasars we identified in the GH-S field were also classified as CT by \citet{brightman2012}, using the same \ch\ data as in this work; the third source was not included in the \citet{brightman2012} sample, but was analysed by \citet{brightman2014} and classified as a CT quasar. Nevertheless, the overlap between our CT quasar candidates and those found by \citet{brightman2012} is small ($\approx20-25$\%, only two sources), as most of our CT candidates are X-ray undetected, while the CT quasars (with intrinsic $L_{\rm X}>10^{44}$~\ergs) identified by \citet{brightman2012} have MIR AGN luminosities below our IR quasar threshold (according to our SED analyses). This is perhaps just a selection effect due to the scatter of the $L_{\rm X}-L_{\rm 6 \mu m}$ relation; however, it might also suggest that the quasar selection approaches in the MIR and X-ray bands are complementary, and tend to identify different source populations (e.g., \citealt{brightman2012}), where the MIR selected CT quasars are brighter at MIR wavelengths and more obscured ({ e.g., \nh${\gtrsim(5-10)\times10^{24}}$~cm$^{-2}$}, as the majority are X-ray undetected) than the X-ray selected CT quasars, possibly because of a larger dust content in the circumnuclear regions. If we combine the number of CT quasars identified in our work over the GH-S field with those found by \citet{brightman2012} within the same area ($\sim$7 sources within the GH-S area, of which 2 are in common with ours), we would obtain a total CT quasar space density of $\Phi=(1.9\pm0.6)\times10^{-5}$~Mpc$^{-3}$. This is in agreement with the predictions by \citet{shi2013} based on a joint model of the X-ray and IR backgrounds. This suggests that the best estimate of the true space density of CT quasars is possibly obtained from a combination of the CT populations identified in the IR and in the X-ray bands.

\subsection{IR quasars and BH-galaxy co-evolution}\label{coev}

According to some evolution scenarios, the most luminous AGN, hosted in the most massive systems, experience a different evolutionary path than that of smaller, less luminous systems. Major mergers are predicted to play a significant role in driving gas into the centre of the most massive galaxies, causing violent episodes of starbursts and rapid SMBH growth (e.g., \citealt{dimatteo2005,hopkins2006a}), especially at high redshifts (e.g., \citealt{hopkins2014}).  On the other hand, secular processes, such as disc instabilities and galaxy bars, are expected to be the main mechanisms driving the gas into the centre of smaller, low-luminosity systems, where the SMBH and the galaxy experience a slower and more steady growth (see also \citealt{alexander2012}, for a review). To place the IR quasars, which are the brightest MIR sources detected in the GH fields, into this evolution scenario, we investigated the host galaxy properties in terms of the SFR and the level of disturbance in the galaxy morphology to detect signs of possible recent interactions/mergers. Our quasars have typically SFRs consistent with those of MS SFGs at redshift $z\approx1-3$, with no indication of unobscured quasars having suppressed SF (i.e., below the star-forming MS) compared to the most obscured ones (Fig. \ref{fig.l6sfr}). We note, however, that amongst the three quasars with the highest SFRs within our sample (SFR$\approx800~M_{\odot}$~yr$^{-1}$), two are CT quasars, one of which is X-ray undetected (\#1), and the remainder is heavily obscured by \nh$\approx5\times10^{23}$~cm$^{-2}$. This suggests that in these sources large amounts of cold gas over the whole galaxy might be partly responsible for the obscuration. However, we do not find any direct correlation between the SFRs and \nh\ over the whole sample, suggesting that in general the X-ray obscuration is probably confined to the nuclear regions and not related to the gas on galaxy scales (e.g., \citealt{rosario2012,rovilos2012}). 

As the morphology of the majority of our sources does not show any strong distortion features ($\sim60$\%), we can infer that mergers are not the main fuelling process amongst the IR quasars (Fig. \ref{fig.morph}). This is in good agreement with the merger/interaction fractions found in less luminous AGN and non-AGN hosting galaxies at similar redshifts by \citet{kocevski2012}, but it is in contrast with the evolution scenario proposed for quasars (e.g., \citealt{dimatteo2005,hopkins2006a}). However the fraction of disturbed/interacting systems ($\sim40$\%) is on average higher than those found for sources at lower redshifts ($\approx15-20$\% at $z<1$; e.g., \citealt{cisternas2011}), in agreement with the observed increase of the galaxy merger rate with redshift (e.g., \citealt{conselice2003, kartaltepe2007}).

Interestingly, the fraction of unobscured/moderately-obscured quasars hosted in disturbed systems is significantly lower ($\approx$20\%) than those in undisturbed galaxies ($\approx$80\%; see Fig. \ref{fig.morph}), while the heavily-obscured quasars are equally found in disturbed and undisturbed galaxies, suggesting that the merger fraction in the most heavily obscured systems might be higher (see also \citealt{kocevski2015}). The difference between the two quasar populations is significant at the 90\% confidence level ($P=0.087$, see Sect. \ref{morph}). If unobscured quasars represent a later stage of the SMBH-galaxy evolution compared to the heavily obscured quasars, the distortion/interaction features due to mergers or galaxy interactions might have faded by the time these quasars are observed as unobscured, as the relaxation time of the galaxy is typically $\sim200-400$~Myr (e.g., \citealt{lotz2010}). This might explain the smaller fraction of disturbed systems we found for the unobscured/moderately-obscured quasars. On the other hand, the most heavily-obscured quasars might still be at a younger stage of evolution after the merger, and therefore their hosts still show the signatures of the recent interactions. We note, however, that we did not find any direct connection between the SFR and X-ray obscuration in our quasars, and thus no evidence for an enhancement of SFR in the heavily obscured quasars in disturbed systems, challenging the predictions from the quasar evolution scenarios. However, as the size of our sample is small, we cannot derive any strong conclusion from our results. Further analyses of larger quasar samples and comparisons with well selected control samples of galaxies are needed to accurately test the SMBH-galaxy evolution scenarios. 

\section{Summary and conclusions}

Through detailed SED decomposition analyses of the 24~$\mu$m detected sources in the GH-N and GH-S fields we identified a sample of 33 luminous IR quasars with AGN MIR luminosity $\nu L_{\rm 6 \mu m}>6\times10^{44}$~\ergs\ at redshift $z\approx1-3$. Considering typical intrinsic $L_{\rm X}-L_{\rm 6 \mu m}$ relations, this luminosity corresponds to an intrinsic X-ray luminosity of $L_{\rm 2-10~keV}>2\times10^{44}$~\ergs, i.e., in the quasar regime. We investigated the X-ray properties of these IR quasars through X-ray spectral analysis for the X-ray detected sources to accurately derive the obscured quasar fraction and constrain the CT quasar population at $z\approx2$. Through our SED analyses, which provide reliable measurements of the intrinsic AGN luminosity and of the host galaxy SFR, and through visual classification of the galaxy disturbance morphology using high resolution HST images, we also investigated the host galaxy properties in relation to the quasar luminosity and the X-ray obscuration to understand the SMBH-galaxy connection in the context of the evolution scenarios. Our main results can be summarised as follows:
\begin{itemize}
\item[--] X-ray spectral analyses of the X-ray detected IR quasars ($\sim$70\% of the sample) show that the majority of the sources (16 out of 24) are obscured by column densities \nh$>10^{22}$~cm$^{-2}$, with more than half of them (9/16) being heavily obscured (\nh$>2\times10^{23}$~cm$^{-2}$). Given that the fraction of obscured AGN decreases at high luminosity (Sect. \ref{frac}){, the number of obscured quasars in our sample is high; however, we find it is} consistent with those expected from an increase of the obscured fraction with redshift ({ after accounting for detection biases,} e.g., \citealt{ueda2014}).
\item[--] Despite being the brightest quasars at MIR wavelengths detected in the GH fields, about 30\% of our IR quasars are not detected in the deepest 2 Ms and 4 Ms \ch\ X-ray data. Their X-ray luminosity upper limits imply a suppression of the X-ray emission by factors of $\sim30-1000$ compared to the intrinsic MIR AGN luminosity. From the comparison of the MIR and observed X-ray luminosities for all the IR quasars in our sample we constrain the fraction of CT quasars to be $\approx{24-48}$\% ({8--16} quasars, { six} of which are robustly identified through X-ray spectral analyses), i.e., up to $\sim${62}\% of all the obscured quasars in the sample (Sect. \ref{obs}).
\item[--] The space density of CT quasars at $z=1-3$ estimated from our sample is $\Phi=({6.7}\pm2.2)\times10^{-6}$ Mpc$^{-3}$, which is in general agreement with previous results for quasars identified in the X-ray or MIR bands \citep{tozzi2006,fiore2009,alexander2008b,brightman2012} as well as with the predictions from the population synthesis models (e.g., \citealt{gilli2007}). However, the small overlap between the CT quasars identified through our analyses and those found in the X-rays within the same fields (e.g., \citealt{brightman2012}) suggests that the space density of CT quasars is likely to be higher than that estimated in this work (or those found in previous studies based on X-ray detected sources), and that the combination of X-ray and MIR selected CT populations possibly provides the best constraints on their true space density in the Universe (Sect. \ref{space}).
\item[--] The SFRs of the IR quasar hosting galaxies are broadly consistent with those of main sequence galaxies at $z\approx1-3$. In general we do not find any direct correlation between the SFR and the quasar luminosity or with the X-ray obscuration, { with only tentative indication for the heavily-obscured quasars (\nh$>2\times10^{23}$~cm$^{-2}$) to have higher SFRs than the less obscured ones (in particular at $z=1.5-2.5$). The limited number of sources in our sample, however, does not allow us to derive any strong conclusion from this result (Sect. \ref{sfr}).}
\item[--] From visual classification of the galaxy morphology to identify signs of interactions/mergers we find that $\approx$40\% of our quasars have a disturbed morphology, while $\approx$60\% do not show any sign of distortions, indicating that mergers are not the main mechanism fuelling the SMBH in luminous quasars. This is consistent with the results found for lower-luminosity AGN (e.g. \citealt{kocevski2012}). However, the interaction/merger fraction seems to be lower for unobscured/moderately-obscured quasars than for heavily-obscured quasars, possibly suggesting that large amounts of gas and dust in the galaxy, as a result of recent merger or galaxy interactions, are partly responsible for the high level of obscuration in these quasars, while the unobscured/moderately-obscured quasars might be at a later evolutionary stage, where the signatures of galaxy interactions/mergers have already faded (Sect. \ref{morph}).  
\end{itemize}
Further progress can be made in the near future to constrain the CT quasar population using the extremely deep 7 Ms \ch\ data recently obtained in the CDF-S (PI. W. Brandt). These data will allow us to confirm the CT nature of the CT quasar candidates we identified in the GH-S field (e.g., providing better spectra and detecting some of the X-ray undetected systems), further validating our analyses and results. Moreover, the advent of the new X-ray and MIR observatories, such as $eROSITA$, $ATHENA$ and the {\it James Webb Space Telescope (JWST)} will provide unprecedentedly large samples of high-$z$ quasars to allow studies of MIR and X-ray selected CT quasars, placing tighter constraints of their space density and on the different properties characterising sources selected in the MIR versus those selected in the X-ray bands.  

\section*{Acknowledgements}
{ We thank the anonymous referee for the useful comments and suggestions that helped improving the manuscript.}
We gratefully acknowledge financial support from the Science and Technology Facilities Council (STFC, ST/L00075X/1; 
 ADM, DMA and CMH) and the Leverhulme Trust (DMA). FEB acknowledges support from CONICYT-Chile (Basal-CATA PFB-06/2007, FONDECYT 1141218, Gemini-CONICYT 32120003, "EMBIGGEN" Anillo ACT1101), and the Ministry of Economy, Development, and Tourism's Millennium Science Initiative through grant IC120009, awarded to The Millennium Institute of Astrophysics, MAS. WNB acknowledges the Chandra X-ray Center grant G04-15130A and NASA ADP grant NNX10AC99G. YQX acknowledges support of the Thousand Young Talents program (KJ2030220004), the 973 Program (2015CB857004), the USTC startup funding (ZC9850290195), the NSFC-11473026, NSFC-11421303, the Strategic Priority Research Program ``The Emergence of Cosmological Structures'' of the Chinese Academy of Sciences (XDB09000000) and the Fundamental Research Funds for the Central Universities (WK3440000001).

\bibliographystyle{mn2e}

\appendix
\section{Infrared Spectral Energy Distributions of the IR quasars}\label{ap1}
\begin{figure*}
\centerline{
\includegraphics[scale=0.83]{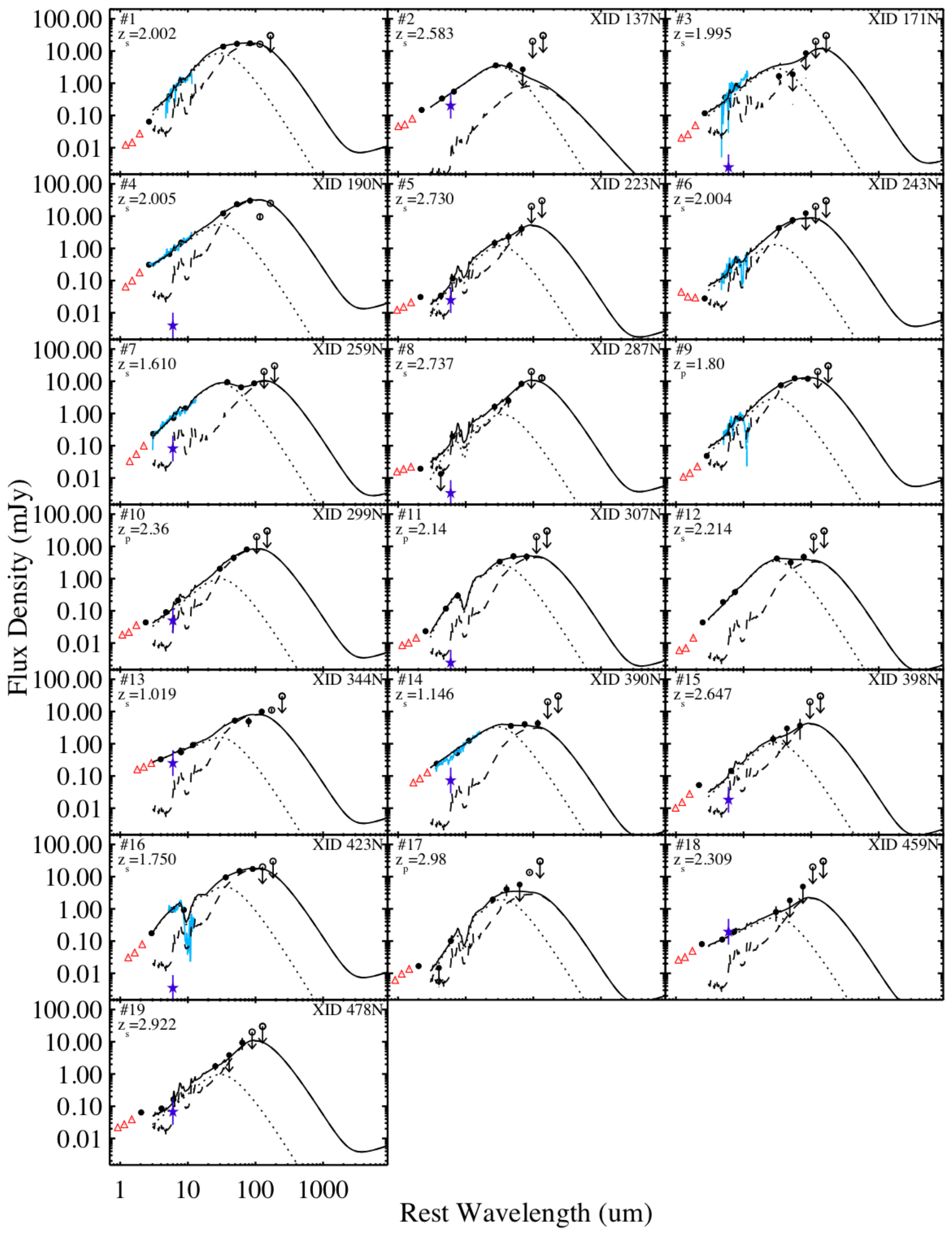}}
\vspace{-0.2cm}
\caption{Spectral energy distribution of the IR quasars in GH-N. Filled circles are the \spz\ 8.0, 16 and 24~$\mu$m and \her\ 100, 160 and 250 $\mu$m used to constrain the SEDs. The \spz-IRAC 3.6, 4.5 and 5.8 $\mu$m data (open triangles) and the \her\ 350 and 500 $\mu$m (open circles) are also shown, although these data are not used in the SED fitting process (Sect. \ref{sed}). For 8 out of 19 IR quasars in GH-N ($\sim$42\%) \spz-IRS spectra are available and they are shown here (cyan line) to demonstrate the agreement with the best-fitting SEDs. The dotted curves and the dashed curves represent the AGN and SFG templates, respectively, while the solid curve is the total best-fitting SED. The purple stars, which are only shown for the X-ray detected IR quasars, represent the 6~$\mu$m flux we would expect from the observed X-ray luminosity adopting the \citet{lutz2004} relation (for XID 243N the star falls below the plotted flux range). On the top left corner of each plot we report the source number and the redshift, while on the top right corner there is the XID from \citet{alexander2003} for the X-ray detected quasars. }\label{fig.sed}
\end{figure*}

\begin{figure*}
\centerline{
\includegraphics[scale=0.83]{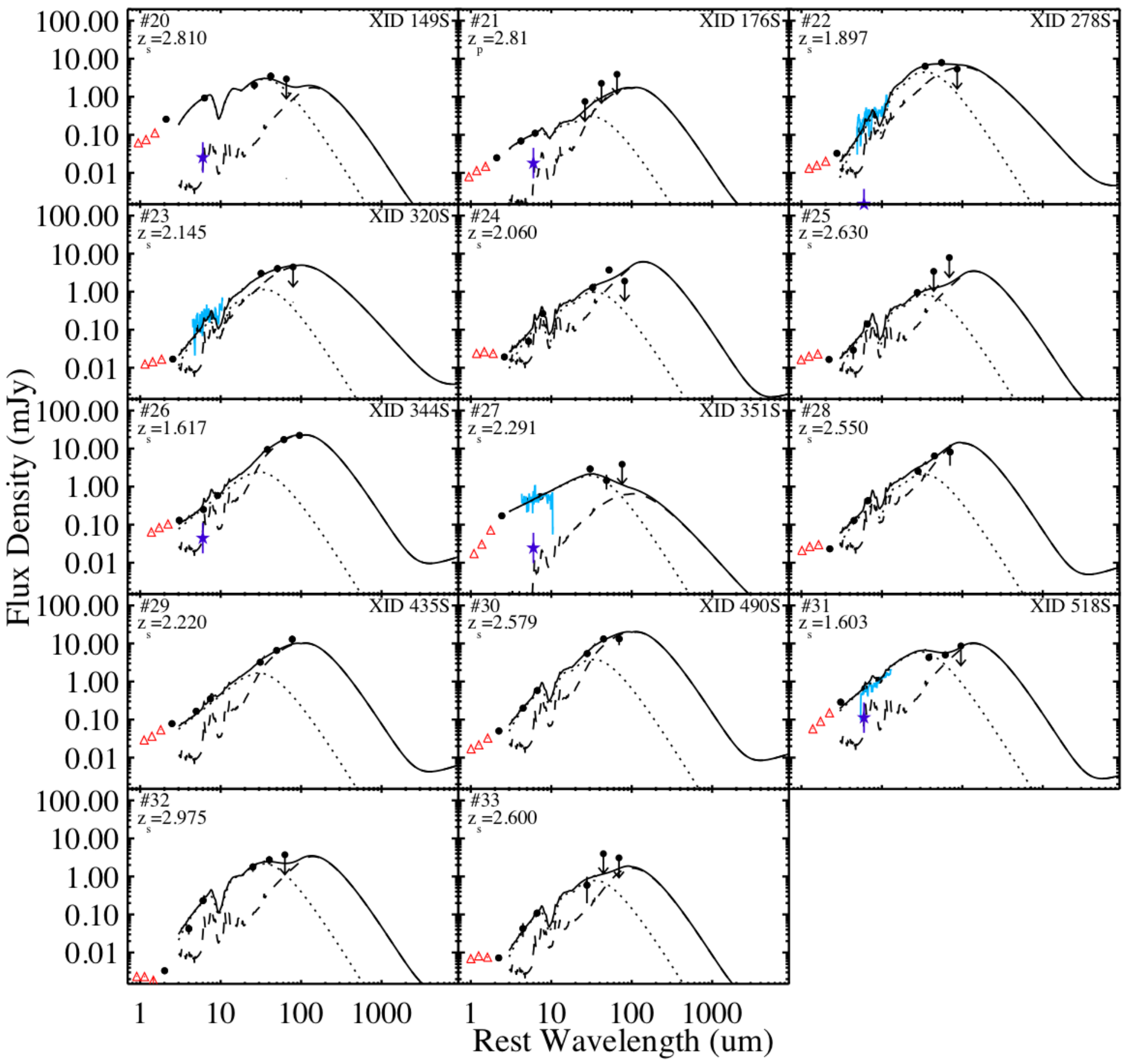}}
\vspace{-0.2cm}
\caption{Spectral energy distribution of the IR quasars in GH-S. Symbols are the same as in Fig. \ref{fig.sed}. The \spz-IRS spectra, which are available for 4 out of the 14 IR quasars in GH-S ($\sim$29\%) are shown (cyan lines) to demonstrate the agreement with the best-fitting SEDs. For XIDs 320S, 435S and 490S, which we robustly identified as CT quasars, the purple  stars, which represent the 6~$\mu$m flux we would expect from the observed X-ray luminosity adopting the \citet{lutz2004} relation, falls below the plotted flux range. On the top left corner of each plot we report the source number and the redshift, while on the top right corner there is the XID from \citet{xue2011} for the X-ray detected quasars.}\label{fig.sed1}
\end{figure*}

\end{document}